\documentclass[a4paper,useAMS,fleqn,usenatbib]{mnras}

\usepackage[T1]{fontenc}
\usepackage{ae,aecompl}

\usepackage{graphicx} 
\usepackage{amsmath}  
\usepackage{amssymb}  
\usepackage{bm}
\usepackage{ulem}

\hypersetup{draft} 

\usepackage{xcolor}

\title[Antlia~2 dwarf]{The hidden giant: discovery of an enormous
  Galactic dwarf satellite in \textit{Gaia} DR2}

\author[G.~Torrealba et al.]{G.~Torrealba,$^{1}$\thanks{E-mail: gtorrealba@asiaa.sinica.edu.tw}
V.~Belokurov,$^{2,3}$
S.~E.~Koposov,$^{4,2}$
T.~S.~Li,$^{5,6}$
M.~G.~Walker$^{4}$
\newauthor
J.~L.~Sanders,$^{2}$
A.~Geringer-Sameth,$^{7}$
D.~B.~Zucker,$^{8,9}$
K.~Kuehn$^{10},$
N.~W.~Evans$^{2},$
\newauthor
W.~Dehnen$^{11, 12}$
\\
$^{1}$ Institute of Astronomy and Astrophysics, Academia Sinica, P.O. Box 23-141, Taipei 10617,Taiwan\\
$^{2}$ Institute of Astronomy, University of Cambridge, Madingley Road, Cambridge CB3 0HA, UK\\
$^{3}$ Center for Computational Astrophysics, Flatiron Institute, 162 5th Avenue, New York, NY 10010, USA\\
$^{4}$ McWilliams Center for Cosmology, Department of Physics, Carnegie Mellon University, 5000 Forbes Avenue, Pittsburgh, PA 15213, USA\\
$^{5}$ Fermi National Accelerator Laboratory, P.O.\ Box 500, Batavia, IL 60510, USA\\
$^{6}$ Kavli Institute for Cosmological Physics, University of Chicago, Chicago, IL 60637, USA\\
$^{7}$ Department of Physics, Imperial College London, Blackett Laboratory, Prince Consort Road, London SW7 2AZ, UK\\
$^{8}$ Department of Physics \& Astronomy, Macquarie University, Sydney, NSW 2109, Australia\\
$^{9}$ Macquarie University Research Centre for Astronomy, Astrophysics and Astrophotonics, Macquarie University, Sydney, NSW 2109, Australia\\
$^{10}$ Australian Astronomical Optics, Macquarie University, North Ryde, NSW 2113, Australia\\
$^{11}$  Department of Physics \& Astronomy, University of Leicester, University Road LE1 7RH Leicester, UK\\
$^{12}$ Sternwarte der Ludwig-Maximilians-Universit\"at, Scheinerstrasse 1, M\"unchen D-81679, Germany
}

\date{Accepted XXX. Received YYY; in original form ZZZ}

\pubyear{2018}

\begin{document}
\label{firstpage}
\pagerange{\pageref{firstpage}--\pageref{lastpage}}
\maketitle


\begin{abstract}
We report the discovery of a Milky Way satellite in the constellation of
Antlia. The Antlia~2 dwarf galaxy is located behind the Galactic disc at a
latitude of $b\sim 11^{\circ}$ and spans $1.26$ degrees, which corresponds to
$\sim2.9$ kpc at its distance of 130 kpc. While similar in spatial extent to the Large
Magellanic Cloud, Antlia~2 is orders of magnitude fainter at $M_V=-9$ mag,
making it by far the lowest surface brightness system known (at $\sim31.9$
mag/arcsec$^2$), $\sim100$ times more diffuse than the so-called ultra diffuse
galaxies. The satellite was identified using a combination of astrometry,
photometry and variability data from \textit{Gaia} Data Release 2, and its
nature confirmed with deep archival DECam imaging, which revealed a
conspicuous BHB signal. We have also obtained follow-up spectroscopy using
AAOmega on the AAT, identifying 159 member stars, and we used them to measure
the dwarf's systemic velocity, $290.9\pm0.5$km/s, its velocity dispersion,
$5.7\pm1.1$ km/s, and mean metallicity, [Fe/H]$=-1.4$. From these properties
we conclude that Antlia~2 inhabits one of the least dense Dark Matter (DM)
halos probed to date. Dynamical modelling and tidal-disruption simulations
suggest that a combination of a cored DM profile and strong tidal stripping
may explain the observed properties of this satellite. The origin of this core
may be consistent with aggressive feedback, or may even require alternatives
to cold dark matter (such as ultra-light bosons).
\end{abstract}




\section{Introduction}

While the population of Galactic low-luminosity dwarf satellites may have been
sculpted by a number of yet-unconstrained physical processes such as cosmic
reionisation \citep[see e.g.][]{Bose2018} and stellar feedback \citep[see
e.g.][]{Fitts2017}, the total number of bright satellites depends strongly
only on the mass of the host galaxy, and thus can be predicted more robustly.
According to, e.g., \citet{Shea2018}, in the Milky Way today there may remain
between 1 and 3 undetected satellites with stellar masses $M_*>10^5M_{\odot}$.
An obvious place where such a satellite might reside is the so-called Zone of
Avoidance \citep[ZOA, see e.g.][]{Shapley1961, KK2000}, a portion of the sky
at low Galactic latitude, affected by elevated dust extinction and a high
density of intervening disc stars. The paucity of Galactic dwarf satellites in
this region was already apparent in the catalogue of \citet{Mateo1998} and has
remained mostly unchanged until the present day \citep[see][]{McConnachie2012}.

Until recently, little had been done to search for Galactic satellites in the
ZOA for obvious reasons. First, the region within $|b|<15^{\circ}$ did not
have contiguous coverage of uniform quality. Second, the foreground disc
populations at these latitudes suffer large amounts of differential reddening,
thus displaying complicated and rapidly varying behaviour in colour-magnitude
space. However, today, thanks to data from ESA's \textit{Gaia} space
observatory \citep[][]{2016A&A...595A...1G}, both of the limiting factors
above can be easily mitigated. For example, \citet{2017MNRAS.470.2702K} used
\textit{Gaia} Data Release 1 \citep[GDR1,][]{2016A&A...595A...2G} to discover
two new star clusters, both with $|b|<10^{\circ}$. They also highlighted
\textit{Gaia}'s potential to detect low-luminosity satellites with surface
brightness levels similar to or fainter than those achieved by much deeper sky
surveys \citep[see also][]{2015MNRAS.453..541A}. As explained in
\citet{2017MNRAS.470.2702K}, what \textit{Gaia} lacks in photometric depth, it
makes up in star/galaxy separation and artefact rejection.
\citet{2018arXiv180506473T} continued to mine the GDR1 data to find an
additional four star clusters all within $10^{\circ}$ degrees of the Galactic
plane. Impressively, the two satellite searches above had to rely solely on
\textit{Gaia} star counts, as no proper motion, colour or variability
information was available as part of GDR1 for the majority of sources.

\begin{table}
    \caption{Properties of the Antlia~2 Dwarf}\label{tab:Properties}
    \centering
    \begin{tabular}{@{}lrrl}
        \hline
        Property               & Antlia~2 Dwarf            & Unit\\
        \hline
        $\alpha({\rm J2000})$    & $143.8868 \pm 0.05$   & deg \\
        $\delta({\rm J2000})$    & $-36.7673 \pm 0.10$   & deg \\
        $l$                      & $264.8955 \pm 0.05$   & deg \\
        $b$                      & $ 11.2479 \pm 0.10$   & deg \\
        $(m-M)$                  & $20.6 \pm 0.11$       & mag\\
        $D_\odot$                & $132 \pm 6$           & kpc\\
        $r_{h}$                  & $1.27 \pm 0.12$       & deg\\
        $r_{h}$                  & $2920 \pm 311$        & pc\\
        1$-$b/a                  & $0.38 \pm 0.08$       & \\
        PA                       & $156 \pm 6$           & deg\\
        $M_V$                    & $-9.03 \pm 0.15$       & mag\\
        $\langle\mu\rangle$(r$<$r$_h$) & $31.9 \pm 0.3$  & mag/arcsec$^{2}$ \\
        $[$Fe/H$]$               & $-1.36 \pm 0.04$      & dex\\
        $\sigma_{[\rm Fe/H]}$               & $0.57\pm0.03$      & dex\\
        $rv_{\rm{helio}}$        & $290.7\pm 0.5$        & km/s\\
        $rv_{\rm{gsr}}$          & $64.3\pm 0.5^{\dagger}$         & km/s\\
        $\sigma_{rv}$            & $5.71\pm1.08$         & km/s\\
        $\mu_{\alpha}\cos\delta$ & $-0.095\pm0.018^{*}$      & mas/yr\\
        $\mu_{\delta}$           & $0.058\pm0.024^{*}$       & mas/yr\\
        $M(r<r_h)$               & $5.5\pm 2.2$          & 10$^7\,M_{\odot}$\\
        $M(r<1.8r_h)$            & $13.7\pm 5.4$         & 10$^7\,M_{\odot}$\\
        $M_\star$                & $8.8\pm 1.2$          & 10$^5\,M_{\odot}$\\
        $M/L_V$                  & $315\pm 130$          & $M_{\odot}/L_{\odot}$\\
        \hline
        \multicolumn{3}{|l|}{$*$Does not consider systematic uncertainties (see text)}\\
        \multicolumn{3}{|l|}{$\dagger$Does not consider LSR uncertainties.}
    \end{tabular}
\end{table}

In this Paper, we use \textit{Gaia} Data Release 2
\citep[GDR2,][]{2018arXiv180409365G} to discover and analyse a new dwarf
satellite galaxy orbiting the Milky Way. The discovery was made at the
Flatiron Gaia Sprint 2018. We are able to exploit not only the positions of
stars detected on-board \textit{Gaia}, but also their colours, proper motions
and parallaxes. Additionally, we take advantage of the large database of
variable stars identified by \textit{Gaia} and supplied as part of GDR2
\citep[see][]{Eyer2018, 2018arXiv180409373H}, in particular, the RR Lyrae
stars \citep[see][]{Clementini2018}. Our search represents the first ``quick
and dirty'' pass through GDR2 data in search of Galactic satellites, and
relies on the fact that all of the currently known Milky Way dwarfs contain at
least one RR Lyrae star \citep[see][]{Sesar2014, 2015AJ....150..160B}. We also
make use of GDR2 parallax measurements to remove the bulk of the foreground
disc population, as suggested in \citet{2015MNRAS.453..541A} and implemented
in e.g. \citet{CloudsArms}.

\begin{figure*}
    \includegraphics[width=\textwidth]{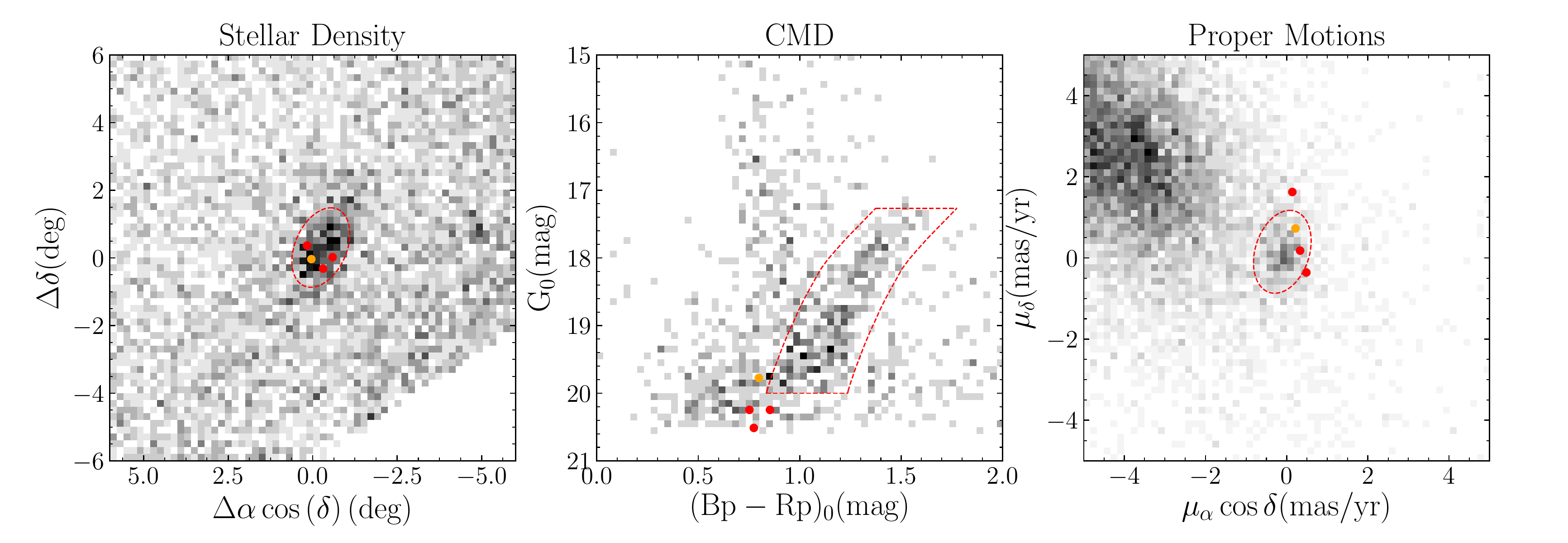}
    \caption{Discovery of Antlia~2 in the \textit{Gaia} DR2 data. {\it Left:}
Proper motion and CMD-filtered stars in a $\sim 100$ deg$^2$ region around
Ant~2. The gigantic elongated overdensity in the center is easily visible once
the proper motion, the CMD and the parallax cuts (see main text) are applied.
Red and orange filled circles, for stars with heliocentric distances larger
than 70 kpc, and between 55 and 70 kpc, respectively, show the position of the
four RRL that we originally used to find Ant~2 (see Section~\ref{s:bhb} and
Figure~\ref{fig:bhbs} for more details). {\it Middle:} CMD of the
\textit{Gaia} DR2 stars within the half-light radius of Ant~2 filtered by
proper motion, featuring an obvious RGB at a distance of $\sim 130$ kpc. {\it
Right:} Stellar PMs within the half-light radius selected using their position
on the CMD, highlighting a clear overdensity around 0. In each panel, red
dashed lines show the selection boundaries used to pick out the likely
satellite members.}
    \label{fig:Gaia}
\end{figure*}

The combined use of \textit{Gaia}'s photometric, astrometric and variability
information allows one to reach levels of surface brightness below those
previously attainable with photometry alone. The extension of the current
Galactic dwarf population to yet fainter systems has been expected, given that
many of the recent satellite discoveries pile up around the edge of
detectability, hovering in the size-luminosity plane around a surface
brightness of $\sim30$ mag arcsec$^{-2}$ \citep[see,
e.g.,][]{2016MNRAS.463..712T}. In other words, it appeared to be only a matter
of time until the ultra-faint galaxy regime would segue into ``stealth''
galaxies, with objects at even lower total luminosities but comparable sizes
\citep[see][]{Bullock2010}. Perhaps even more surprising is the recent
detection of a galaxy - the Crater~2 dwarf - with an extremely low
surface-brightness, but at a total luminosity close to the classical dwarf
regime, i.e. $L\sim 10^5 L_{\odot}$ \citep[see][]{2016MNRAS.459.2370T}. Cra~2
occupies a poorly-explored region of structural parameter space, where
ordinary stellar masses meet extraordinarily large sizes, resulting in
record-breaking low surface brightness levels ($\sim30.6$ mag arcsec$^2$) - a
regime not predicted to be populated by earlier extrapolations
\citep[e.g.][]{Bullock2010}. Stranger still, Cra~2 appears to be not only one
of the largest Milky Way dwarfs, but also one of its coldest \citep[in terms
of the stellar velocity dispersion, see][]{2017ApJ...839...20C}. Of the
plausible mechanisms capable of dialing down both the satellite's surface
brightness and velocity dispersion, tidal stripping immediately comes to mind
\citep[see e.g.][]{Penarrubia2008}. But, as shown by \citet{Sanders2018}, it
is rather difficult to produce a diffuse and cold system such as Cra 2 only
via the tidal stripping of a stellar system embedded in a cuspy \citep[see,
e.g.][]{Dubinski1991,NFW1996} dark matter halo.

As \citet{Sanders2018} convincingly demonstrate, however, if Cra 2 were
embedded in a cored \citep[i.e., shallower inner density -- see,
e.g.][]{Moore1994,Navarro1996} dark matter (DM) halo, reproducing its present
structural and kinematic properties would be much easier. Such cores can
naturally arise if the physics of the DM particle is altered \citep[see,
e.g.][]{Spergel2000,Hogan2000,Hu2000,Peebles2000}, but even within Cold DM
cosmology the inner density profiles of galaxy-hosting halos can be
substantially flattened via strong stellar feedback
\citep[e.g.][]{Elzant2001,Gnedin2004,Read2005,Mashchenko2008,Pontzen2012}.
While the study of the effects of supernova feedback on the structure of
galaxies currently remains firmly in the realm of ``sub-grid'' physics, many
simulations show that the changes induced are not limited to the dwarf's
central regions. Powerful bursty gaseous outflows have been shown to be able
to ``drag'' many of the constituent stars to much larger radii overall, thus
creating noticeably diffuse dwarf galaxies \citep[see,
e.g.][]{Elbadry2016,Dicintio2017,Chan2018}.

This Paper is organized as follows. Section \ref{sec:data} gives the details
of the search algorithm and archival imaging processing; it also describes the
modelling of the structural properties of the system and the estimates of its
distance. Section~\ref{sec:spec} presents the analysis of the spectroscopic
follow-up as well as the details of the kinematic modelling.
Section~\ref{sec:discuss} compares the new satellite to the population of
previously known Milky Way dwarfs, and gives an interpretation of its DM
properties. Concluding remarks can be found in Section~\ref{sec:conc}.

\section{The hidden giant}\label{sec:data}

\subsection{Discovery in \textit{Gaia} DR2}

\textit{Gaia} DR2 boasts many unique properties that allow one to study the
outskirts of the Milky Way as never before. Perhaps the most valuable of these
is the wealth of high-quality all-sky proper motion (PM) information.
\textit{Gaia}'s astrometry makes it possible to filter out nearby
contaminating populations, revealing the distant halo behind them. Halo
studies are further boosted by the use of \textit{Gaia}'s variable star data,
specifically the RR Lyrae (RRL) catalogue \citep[see][for
details]{2018arXiv180409373H}, which provides precise distances out to (and
slightly beyond) $\sim$100\, kpc. RRLs are the archetypal old, metal-poor
stars, and hence a perfect tracer of the Milky Way's halo, including the dwarf
satellite galaxies residing in it.  Indeed, all but one dSph \citep[Carina
III, see][ for further discussion]{2018MNRAS.475.5085T} that have been studied
so far contain at least one RRL \citep{2015AJ....150..160B}. This makes
searches for stellar systems co-distant with RRLs a plausible means to probe
for low surface brightness Milky Way halo sub-structure \citep[see, e.g.
][]{Sesar2014,2015AJ....150..160B}.

\begin{figure*}
    \includegraphics[width=\textwidth]{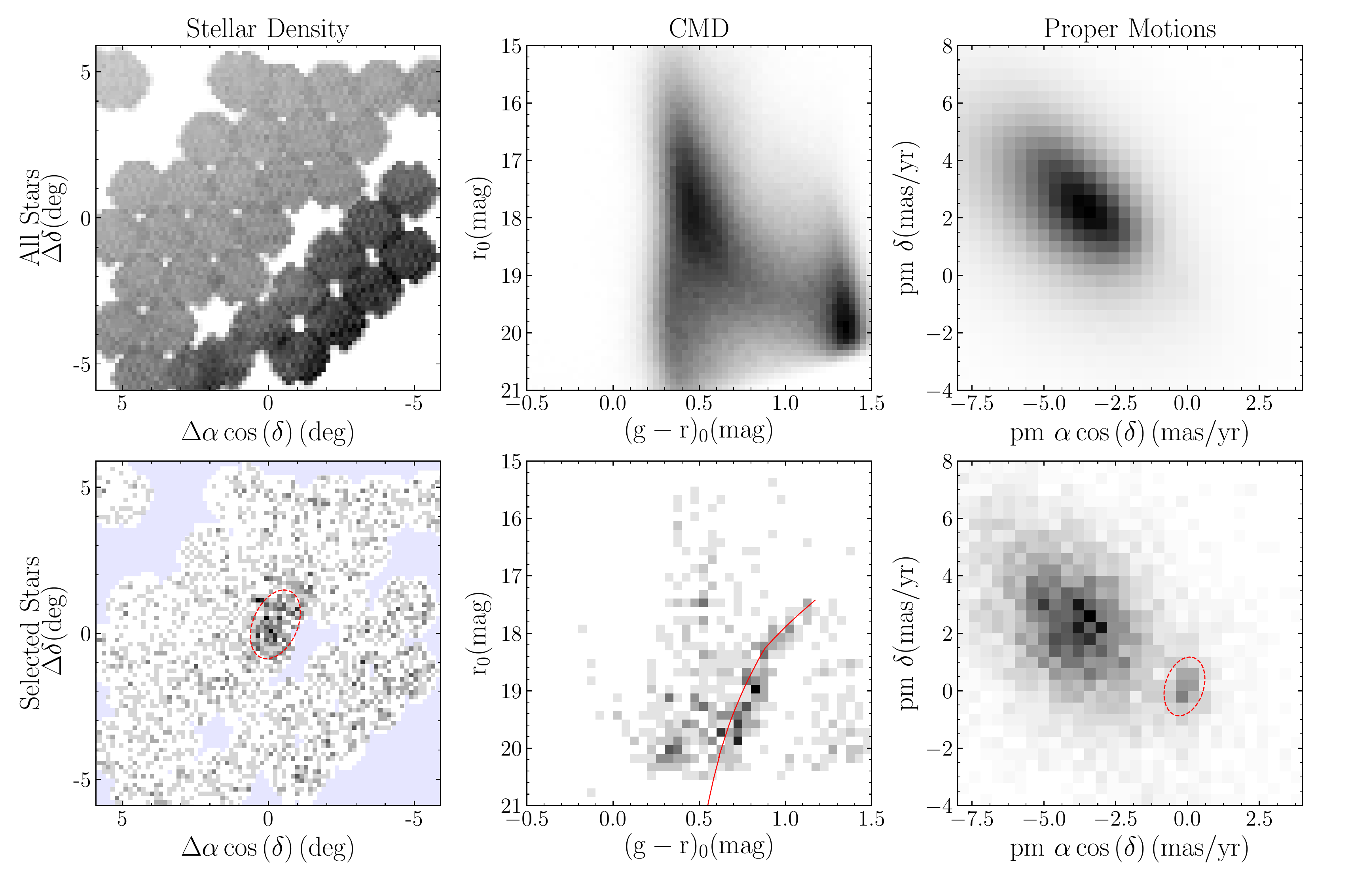}
    \caption{DECam view of Antlia 2. Stars in a $\sim100$ deg$^2$ region
centred on the dwarf with available PMs from GDR2 are shown. The top row gives
the properties of all stars in this portion of the sky, while the bottom
panels illustrate the properties of the likely Ant~2 members. {\it Left:}
Distribution of the stellar density on the sky. The red ellipse corresponds to
the half-light radius of the best-fit model ($r_h=1.26$\degr), and marks the
boundary used for the spatial selection. The blue shaded regions in the lower
panel indicate the areas without DECam data. {\it Middle:} Hess diagram, i.e.
stellar density in the CMD. The red line corresponds to the best-fit isochrone
with $\log$ age = 10 and $[$Fe/H$]=-1.5$. {\it Right:} Stellar density in
proper motion space. The red ellipse marks the PM selection boundary. Being
very close to the Galactic disc ($b\sim11\degr)$, the region around Ant~2 is
heavily dominated by the MW foreground. Nonetheless, after applying all of the
selection cuts (including the parallax, see main text), Ant~2's signal appears
conspicuous in all three parameter spaces.}
    \label{fig:Cuts}
\end{figure*}

In this work, we combine the use of both \textit{Gaia}'s astrometry and its
RRL catalogues to look for previously unknown MW satellites.  We use a clean
sample of RRLs from the \textit{gaiadr2.vari\_rrlyrae} table provided by
\textit{Gaia} DR2, and look for overdensities of stars with the same proper
motions as the RRL considered. Specifically, we first estimate the RRL
distance modulus as

\begin{equation}
  D_h=\left<G\right>-3.1\frac{A_{G}}{A_{V}}E(B-V)-0.5,
\end{equation}

\noindent where $\left<G\right>$ is the intensity-averaged $G$ magnitude,
E(B-V) is taken from the \citet{SFD} extinction map, and $A_G/A_V=0.859$ is
the extinction coefficient for the \textit{Gaia} $G$ band
\citep{2018MNRAS.481.3442M}\footnotemark.  For simplicity we assumed an
absolute magnitude for the RRL of 0.5 \citep[but see][]{Iorio2018}. Then, we
cleaned up the RRL sample by removing stars with
\textit{astrometric\_excess\_noise} larger than 1, and reduced the sample to
search only around stars that have $D_h>50$ and that are at least 15 degrees
away from the LMC and SMC. The stars selected for the overdensity search were
taken in a 2 degree radius from the central RRL. Only stars with PMs
consistent - within the uncertainties - with the central RRL PM were
considered. Additionally, we removed stars with low heliocentric distances by
applying a cut on parallax of $\varpi>0.5$.

\footnotetext{ The coefficients for the $B_P$ and $R_P$ used to create figure
\ref{fig:Gaia} are $A_{B_P}/A_V=1.068$ and
$A_{R_P}/A_V=0.652$\citep{2018MNRAS.481.3442M}. Note, however, that reddening
corrections have a typical uncertainty of around 10\%, owing to both the
scatter in the fit of the extinction coefficient, and the variability of $R_V$
\citep[see Appendix B of][for further details.]{SFD}.}

Specifically, the overdensity search was performed as follows. We counted the
number of the previously selected stars within circular apertures ranging in
radius from 1\arcmin~to 30\arcmin~(from the central RRL), and compared these
to the foreground, which was estimated in the area between 1 and 2 degrees
away from the RRL star. If any of the samples exceeded by more than 2$\sigma$
the expected foreground number, we flagged the trial RRL as a possible tracer
of a stellar system. By plotting the flagged RRL in the sky - together with
the known satellite galaxies and the globular clusters - we immediately
noticed that three flagged RRL with distances between 55 and 90 kpc were
bunching up in a small region of the sky where no known stellar system was
present. A closer inspection revealed that the RRLs shared the same PM;
moreover the PM-filtered stars in the region had a conspicuous signal both in
the CMD and on the sky, as seen in Figure \ref{fig:Gaia}. More precisely, the
left panel of the Figure shows the spatial distribution of the stars selected
using the PM and the CMD cuts. Here a large stellar overdensity spanning more
than 1 degree on the sky is visible. The middle panel gives the CMD density of
stars within the half-light radius (red dashed line in left panel, see Sec.
\ref{sec:photmod} for its definition), and after applying the PM cut. A broad
Red Giant Branch (RGB) at a distance of $\gtrsim100$kpc can be easily
discerned. The red polygon indicates the CMD mask used for the selection,
which was empirically defined based on the RGB feature. Finally, the right
panel of the Figure demonstrates the PM density of stellar sources within the
half-light radius and inside the CMD mask shown in the middle panel. A tight
over-density is noticeable around $\mu_{\alpha}\sim 0$ and
$\mu_{\delta}\sim0$. The red ellipse outlines the PM selection boundary, as
defined in Section~\ref{sec:photmod}. Based only on the approximate distance,
the size and the breadth of the RGB we can safely conclude that the newly
found object is a dwarf galaxy satellite of the Milky Way. This hypothesis is
further tested and confirmed below. As the dwarf is discovered in the
constellation of Antlia (or the Pump), we have given it the name Antlia~2 (or
Ant~2). Note that Ant~2's neighbour on the sky, the previously-found Antlia
dwarf, is a transition-type dwarf (i.e. a galaxy with properties similar to
both dwarf spheroidals and dwarf irregulars) on the outskirts of the Local
Group \citep[i.e. beyond 1 Mpc, see][]{1997AJ....114..996W}.

\begin{figure*}
    \includegraphics[width=\textwidth]{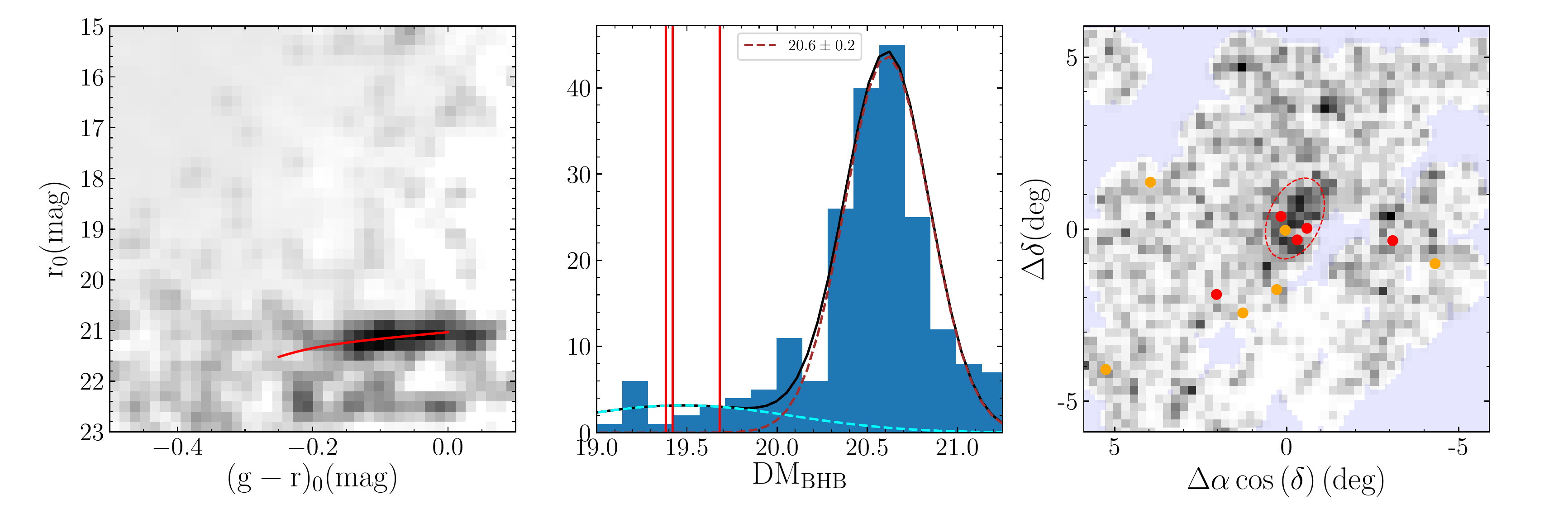}
    \caption{BHBs and RR Lyrae stars in Antlia~2. {\it Left:} DECam Hess
difference (difference of CMD densities) of the stars within the Ant~2's
half-light radius and the foreground, zoomed in on the BHB region. The red
line shows the BHB ridge-line as given in \citet{Deason2011} at the best fit
distance modulus. {\it Middle:} Distribution of the distance moduli of the selected BHB
candidates, along with the best-fit two-Gaussian plus a flat background model.
The vertical lines mark the distance moduli of the four RRL stars within the
half-light radius of the satellite. {\it Right:} On-sky density distribution
of the BHB candidates. RRL stars with heliocentric distances larger than 70
kpc are shown in red, stars with distances between 55 and 70 kpc are shown in
orange. }
    \label{fig:bhbs}
\end{figure*}

\subsection{Photometric modeling}\label{sec:photmod}

\subsubsection{Deep DECam imaging data}

To better characterise the new object, we checked whether any deeper
photometric data were available. In this region, photometry with partial
coverage was found in the NOAO source catalogue
\citep[NSC,][]{2018arXiv180101885N}; additionally several unprocessed images
were available from the NOAO archive. We searched for DECam data in the g and
r bands in a 100 square degree region around $(ra,dec)=144.1558,-37.07509$,
and retrieved the \textit{instcal} and weight frames provided by NOAO. Most of
the area is covered with images from two Programs, namely 2017A-0260 (the
BLISS survey) and 2015A-0609, but we also downloaded images from Programs
2013B-0440 and 2014B-0440, although these latter two only added 3 fields in
total in the outskirts of the region.

Photometry was carried out using the standard SExtractor+PSFex combination
\citep[see, e.g.][for a similar approach]{Koposov2015}. We kept the
configuration standard, except for a smaller detection threshold, which was
set to 1, and a more flexible deblending threshold. The photometric zero point
was calibrated against the ninth APASS data release
\citep{2016yCat.2336....0H} as the median magnitude offset on a per-chip
basis. If fewer than 10 stars were available on a chip, then the field median
offset was used instead. We finalised the calibration with a global
correction. This procedure gave a photometric zero-point precision of 0.078 in
$g$ band and 0.075 in $r$ band.  To generate the final band-merged source
catalogues, we first removed duplicates using a matching radius of 1\arcsec,
and then cross-matched the $g$ and $r$ band lists with the same matching
radius, only keeping objects that had measurements in both bands. We also
cross-matched the resulting catalogue with the \textit{Gaia} DR2 source list -
also using a 1\arcsec\ radius - to complement the DECam photometry with PM
information where available. The final catalogue covers $\sim 88$ square
degrees, of which $\sim 65$ come from Program 2017A-0260 with a limiting
magnitude of $\sim 23.2$. The remaining $23\,\rm{deg}^2$ are from 2015A-0609
and have a limiting magnitude of $22.2$ (both before extinction correction).
The extinction correction was done using the dust maps from \citet{SFD} and
the extinction coefficients from \citet{Schlafly2011}. Note that Antlia~2 is
in a region of high extinction, with $\sim0.6$ mag of extinction in $g$ and
$\sim0.4$ mag in $r$, which adds $\sim0.05$ mag to the uncertainty if one
considers the 10\% uncertainty on the reddening correction \citep{SFD}.
Throughout the paper we usually refer to the extinction-corrected magnitudes,
which are labeled with the subscript 0. Finally, likely stars are separated
from the likely galaxies by removing objects with \textit{SPREAD\_MODEL}
greater than $0.003$+\textit{SPREADERR\_MODEL} in both $g$ and $r$ bands.

Figure \ref{fig:Cuts} shows the area of the sky around Ant~2 in the archival
DECam data. From left to right, we present the stellar density distribution on
the sky, the density of stars in CMD space, and density of stars in PM space.
The top panels show all stars in our DECam catalogues, while the bottom panels
show only the selected candidate Ant~2 member stars. Note that, although we
have photometric catalogues that reach down to $r_0\sim23.2$, we only use
stars cross-matched to the GDR2 catalogues due to the need for PMs for our
analysis. The GDR2's PMs are only available - with high completeness - down to
20.4 in $g_0$, and to 20 in $r_0$. As the top row of the Figure demonstrates,
the object is essentially invisible when no filters are applied.  Given the
complex, overpowering stellar foreground population, we decided to
characterise the object independently in the three parameter spaces.

\begin{figure}
     \includegraphics[width=0.5\textwidth]{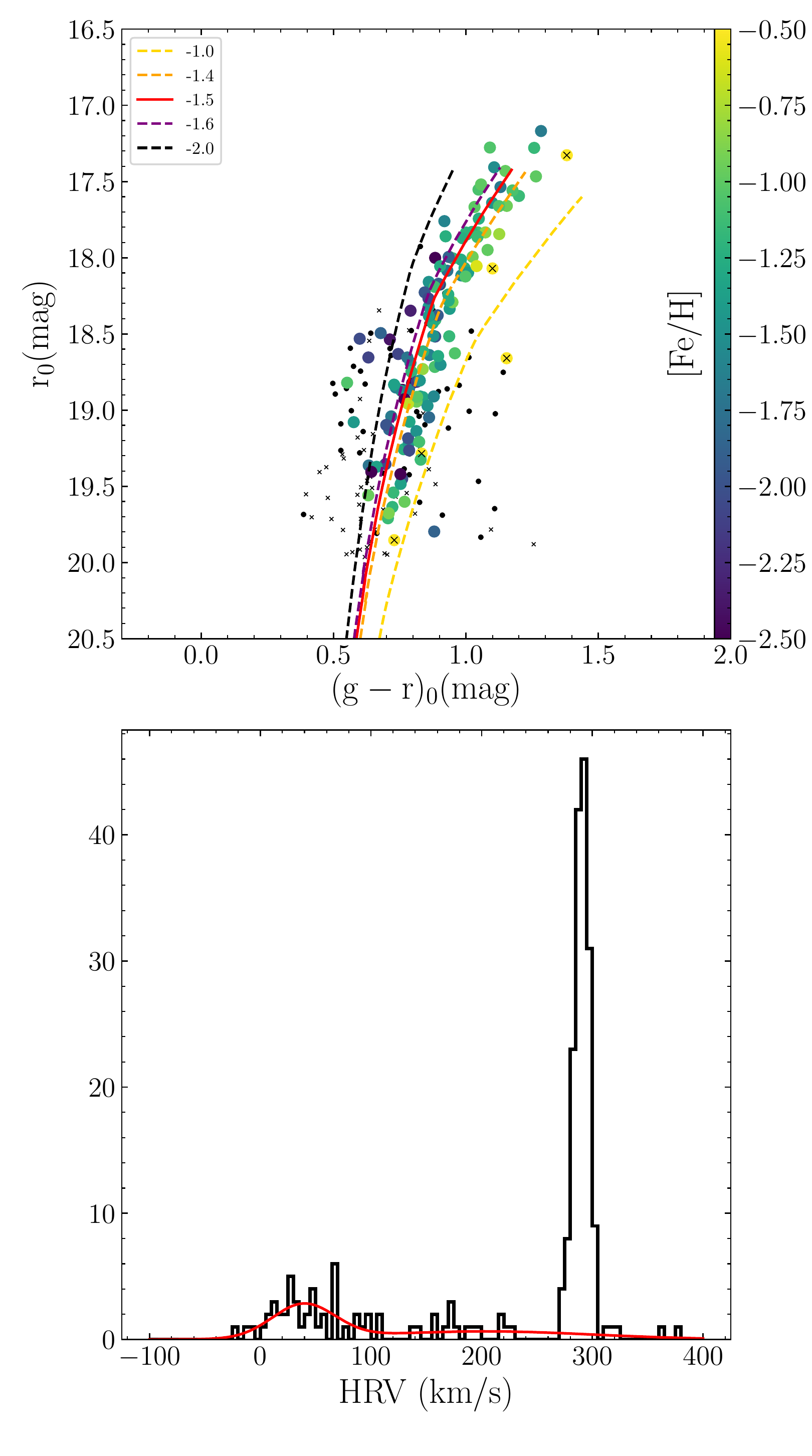}
     \caption{{\it Top:} CMD of the stars with spectroscopic measurements
within the DECam data footprint. Ant~2 stars with accurate velocity
measurements are colour-coded according to their spectroscopic metallicity.
Small black dots are stars with velocities inconsistent with Ant~2 membership
and small black crosses are stars without good velocity measurements. The red
line corresponds to the best-fit isochrone from the photometric modeling,
while the dashed lines show isochrones with the same age, but different
metallicity values. Reassuringly, there is a good correlation between the
spectroscopic metallicity of each star and the general ($(g-r)_0$ colour)
trend marked by the different isochrones. The large spread in the
spectroscopic metallicity of Ant~2 stars appears to be consistent with the
large width of the RGB. {\it Bottom:} Distribution of heliocentric radial
velocities of the targeted stars. Only stars with accurate velocity
measurements are shown (uncertainties less than 10\, km\,s$^{-1}$, residual
kurtosis and skewness less than 1). The red curve shows the best-fit
foreground model, consisting of two Gaussian distributions.}
     \label{fig:specres}
\end{figure}

We start by modelling the distribution of the dwarf's PMs, since in this space
the satellite's signal can be most easily differentiated from that of the
foreground. To proceed, we apply the spatial and the CMD cuts based on the
signals seen in the bottom row of Figure \ref{fig:Cuts}. Then we model the
resulting PM distribution as a mixture of three Gaussians, two representing
the foreground and one for Ant~2 itself. The bottom right panel of Figure
\ref{fig:Cuts} gives the density distribution of the CMD+spatially selected
stars along with a contour corresponding to the best fit-Gaussian shown in a
red dashed line.  The resulting Gaussian profile, centred at
$\left(\rm{pm}\,\alpha\cos\left(\delta\right),\rm{pm}\,\delta\right)=(-0.1,0.15)$,
provides a good description of the PM ``blob'' seen in the Figure. The red
dashed contour shown in the right panel of Figure \ref{fig:Gaia} and in the
bottom right panel of Figure \ref{fig:Cuts} is the mask we apply to select
stars that belong to Antlia~2 based on PMs.

Next, we analyse the observed stellar spatial distribution using the full 88
square degrees of DECam imaging available. We first apply the PM and the CMD
cuts, and then model the resulting distribution as a mixture of a planar
foreground and a Plummer profile, following the same procedure as described
in, e.g., \citet{2018arXiv180506473T}. In order to obtain robust and useful
uncertainties for the parameters of our spatial model, we sample the
likelihood using the ensemble sampler \citep{GoodmanWeare2010} implemented in
the \textit{emcee} package \citep{ForemanMackey2013}. We chose flat priors for
all parameters except for the dwarf's size, which uses the Jeffreys prior,
i.e., 1/$a$, where $a$ is the size parameter. The best-fit parameters and
their uncertainties are estimated from the marginalized posterior
distributions corresponding to the $15.9\%$, $50\%$, and $84.1\%$ percentiles.
The half-light radius of the best-fit model is shown as the red ellipse in the
bottom left panel of Figure \ref{fig:Cuts}, corresponding to a whopping
$r_h=1.26\pm0.12$ deg. The Ant~2 dwarf's angular extent is similar to that of
the SMC \citep[1.25 deg,][]{2016MNRAS.459.2370T}, with only the LMC \citep[2.5
deg,][]{2016MNRAS.459.2370T} and Sagittarius \citep[5.7
deg,][]{McConnachie2012} being larger in the sky. Curiously, this is almost
three times larger than the next largest known satellites, Cra2 and BooIII,
both with $r_h\sim0.5$ deg
\citep[][]{2016MNRAS.459.2370T,2009ApJ...693.1118G}. Located at $\sim130$ kpc
(see sec. \ref{s:bhb} for details), this angular extent translates to
$\sim2.9\pm0.3$ kpc, which is equal in size to the LMC! A summary of the
relevant physical parameters measured above is presented in
Table~\ref{tab:Properties}.

\begin{figure*}
     \includegraphics[width=\textwidth]{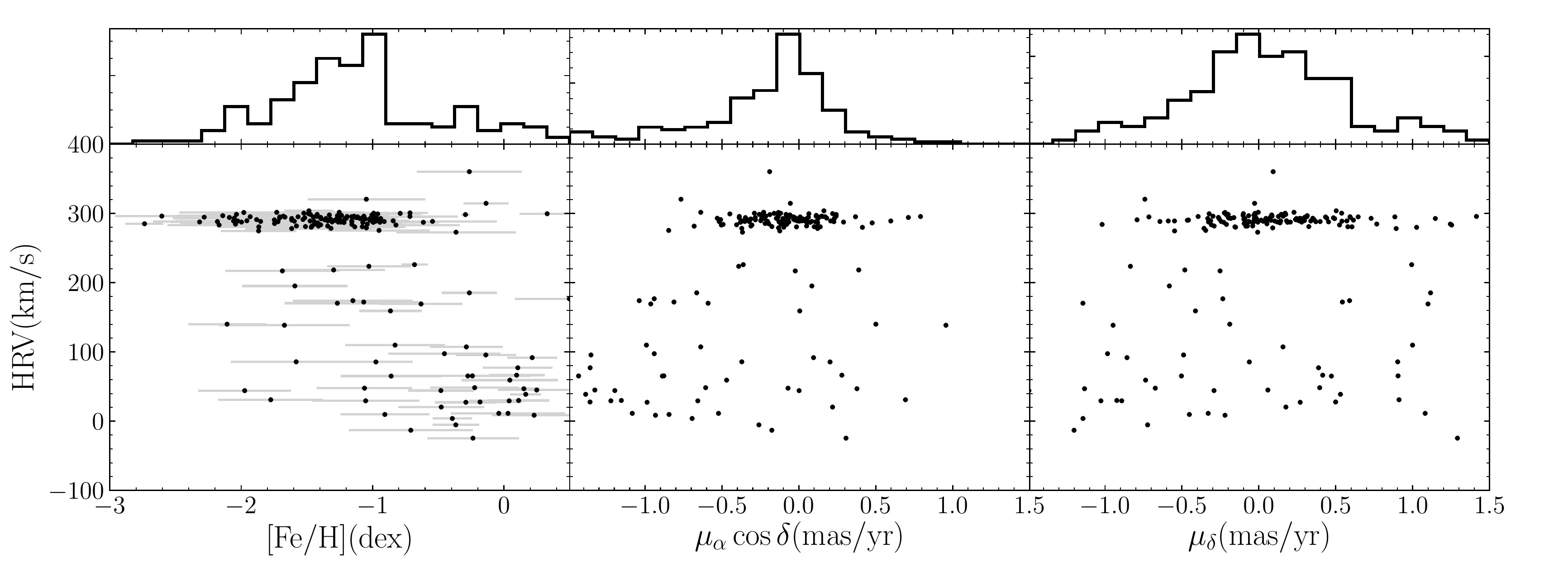}
     \caption{Heliocentric radial velocities of Ant~2 targets. {\it Left:} HRV
vs spectroscopic metallicity. Only stars with good RVs (see
Fig.~\ref{fig:specres}) and small metallicity errors $\sigma_{[Fe/H]}$<0.5 are
shown. Ant~2 members stand out clearly as low-metallicity stars, while a small
number of foreground contaminants are mostly high-metallicity disc stars.
There is also a handful of halo stars. {\it Middle/Right: } HRV vs proper
motion in Right Ascension (middle) and Declination (right). Since only stars
within $\pm 1.5$ mas\,yr$^{-1}$ were spectroscopically targeted, the proper
motion range shown is truncated. Note however that Ant~2's proper motion is
clearly distinct from the bulk of the foreground contaminants and concentrates
towards $\mu\approx 0$. This can also be seen in the corresponding 1D
histograms shown above each panel.}
     \label{fig:rv_feh}
\end{figure*}

Finally, we model the stellar distribution in CMD space using Padova
isochrones \citep{Bressan2012} for the satellite's stars and an empirically
estimated foreground. In line with previous steps, we first apply a spatial
cut, using the best-fit structural model, as well as a PM cut based on the
constraint derived above. Additionally, we fix the distance modulus to the
value obtained by fitting the satellite BHB candidates (see section
\ref{s:bhb} for details); more precisely we use $m-M=20.6$, and we set up a
magnitude limit of 20 in $r_0$ and 20.4 in $g_0$, corresponding to the range
within which all stars have their PMs measured by \textit{Gaia}. The isochrone
models are built on a colour-magnitude grid by convolving the expected number
of stars along the isochrone with the photometric uncertainties. We also
convolve the maps with a Gaussian with $\sigma=0.2$ mag along the magnitude
component to account for the observed spread in distance modulus, which
corresponds to the full width of the BHB sequence (see the middle panel of
Fig.~\ref{fig:bhbs}). The modeling is performed only between
$0.5<(g-r)_0<1.5$, since this is the region where the RGB, the only easily
discernible CMD feature, is located. To create the foreground model, we make a
density map of the stars in the same part of the sky, but removing the stars
with the PM of Ant~2. Finally, we pick isochrones on a grid of logarithmic
ages between $9.6$ and $10.1$ and metallicities between $-2.1$ and $-0.8$ and
measure the likelihood of the data given each isochrone. For each isochrone,
the only parameter we fit is the ratio of the foreground stars to the
satellite's. The best-fit model obtained is that with $\log$ (age) = 10.0 and
$[$Fe/H$]=-1.5$. The best-fit isochrone along with the PM and spatial filtered
CMD is shown in the middle panel of Figure \ref{fig:Cuts}. Note that the good
fit of the RGB at the given distance modulus provides an independent
confirmation of the distance to Ant~2. Using the above CMD model, specifically
the ratio of satellite members to background/foreground, and the spatial model
to account for chip-gaps, we can infer the total number of Ant~2's stars above
the limiting magnitude to be $N=246\pm30$.  Given that the catalogue is close
to 100\% completeness for $g_0<20.4$, we can combine the total number of stars
with the best-fit isochrone, which assumes a Chabrier IMF
\citep{Chabrier2003}, to estimate the absolute magnitude of Ant~2 as
$M_V=-9.03\pm0.15$, which is equivalent to a stellar mass of
$M_\star=2.5\times10^{0.4\left(4.83-M_v\right)}=(8.8\pm1.2)\times10^5
M_\odot$.

\subsection{Horizontal Branch and distance}\label{s:bhb}

If no PM cut is applied, it is very difficult to see the RGB feature in the
CMD, but it is still possible to see a strong BHB sequence at $r_0\sim21.2$. 
This is illustrated in the left panel of Figure~\ref{fig:bhbs}, in which we
show the blue part of the differential Hess diagram between the stars within
the half-light radius of Ant~2 and those in the foreground. The sample shown
comes from the region where the limiting magnitude in the $r$ band is 23.2.
The red line indicates the BHB ridge-line from \citet{Deason2011} at the
distance modulus of 20.6. This is the best-fit value obtained by measuring the
distance modulus of all stars within the dashed red box, assuming they are
drawn from the above ridge-line. The distribution of the observed distance
moduli is shown in the middle panel of the Figure, along with a two-Gaussian
model, where one component describes the peak associated with Antlia 2 BHBs,
and the other models the foreground. The main peak at $\sim20.6$ is well fit
by the Gaussian with a width of 0.2 magnitudes. Note however that the formal
uncertainty on the center of the Gaussian is only 0.02. Nevertheless there is
a systematic uncertainty of $\sim 0.1$\,mag in the absolute magnitude of the
BHB ridge line itself \citep[][Fig. 4]{Deason2011} which sets the uncertainty
in our DM measurement to 0.1.

This translates into a distance of $132 \pm 6$ kpc. The red vertical lines in
the middle panel of Figure~\ref{fig:bhbs} show the distance moduli of the
three RRL originally found around Ant~2. Clearly, these variable stars --
while located far in the halo -- appear to be positioned well in front of the
dwarf along the line of sight. We speculate that this handful of RR Lyrae
detected by \textit{Gaia} may be on the near side of an extended cloud of
tidal debris (see Section~\ref{sec:discuss} for details) emanating from the
dwarf. Note, however, that all three RR Lyrae lie close to the limiting
magnitude of Gaia, therefore their median flux estimate may be biased high
and, correspondingly, their distances biased low. Given its luminosity, Ant 2
is likely to host many tens of RR Lyrae, similar to, e.g., its close analogue,
Crater 2 \citep[see][]{Joo2018,2018MNRAS.479.4279M}. At the distance of the
main body of the dwarf, RR Lyrae would be too faint for \textit{Gaia} but
should be detectable with deeper follow-up imaging. The distribution of the
DECaM BHBs, along with the RRLs with distances larger than 55 kpc, is shown in
the right panel of Figure~\ref{fig:bhbs}. An obvious BHB overdensity with a
shape very similar to that of the RGB stars is visible at the position of
Antlia~2, further confirming that the BHB and the RGB features are correlated.

\begin{figure}
     \includegraphics[width=0.5\textwidth]{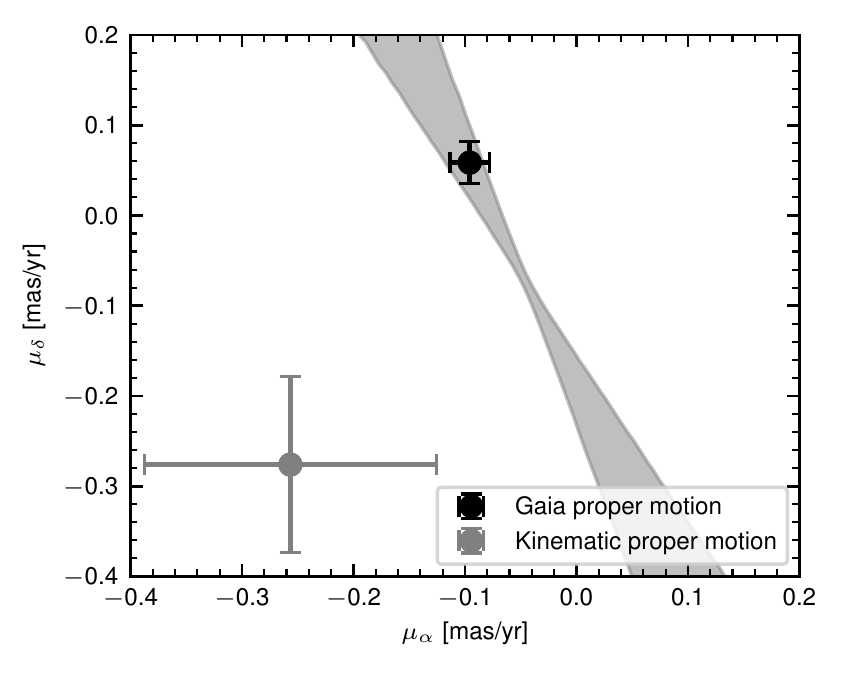}
     \caption{Ant~2's proper motion measurement using two different methods
(see main text). The black filled circle  shows the proper motion of Ant~2's
centre as measured using the \textit{Gaia} DR2 data. The grey filled circle
shows the proper motion inferred from the radial velocity gradient
\citep{Walker2006}. The grey shaded region shows expected proper motions if
Ant~2 moves in the direction indicated by  the elongation of Ant~2's
iso-density contours, as measured in Section \ref{sec:photmod}. The width of
the region is driven by uncertainties in Ant~2's distance and iso-density
position angle. Note that the \textit{Gaia} proper motion is well aligned with
the elongation, suggesting that the elongation may be of tidal nature. The
fact that the kinematic proper motions are pointing in a slightly different
direction suggests that the dwarf's internal kinematics may be affected by
either intrinsic rotation or Galactic tides.}
     \label{fig:pms}
\end{figure}
\begin{figure}
    \includegraphics{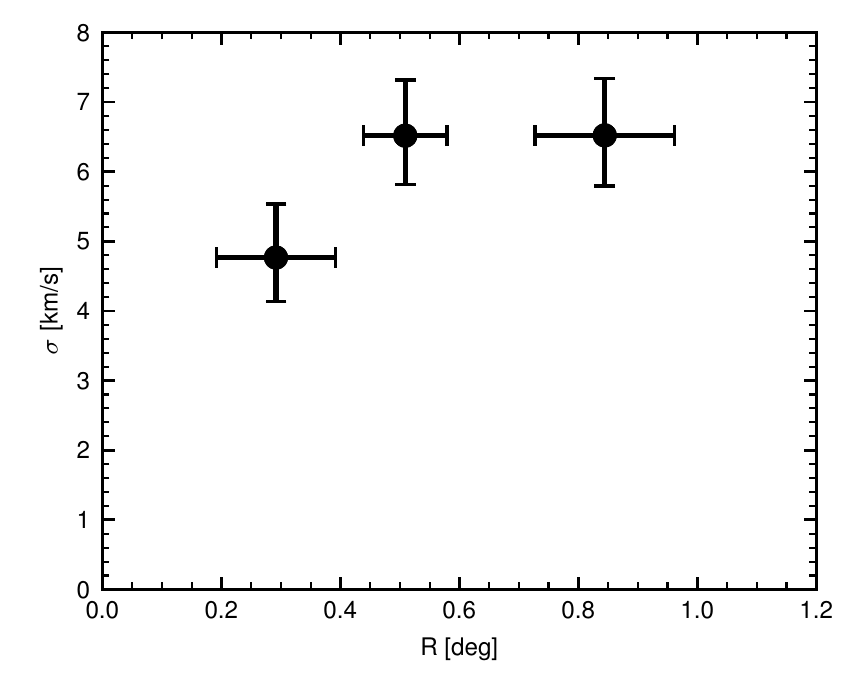}
    \caption{Line-of-sight velocity dispersion measured as a function of radial
distance from Ant~2's centre. Each bin contains roughly equal numbers of
stars. Error bars correspond to the 16\%-84\% percentiles of the measurements.
There is a non-statistically significant hint of a velocity dispersion
decrease close to the center.}
    \label{fig:veldisp_grad}
\end{figure}
\begin{figure*}
     \includegraphics[width=\textwidth]{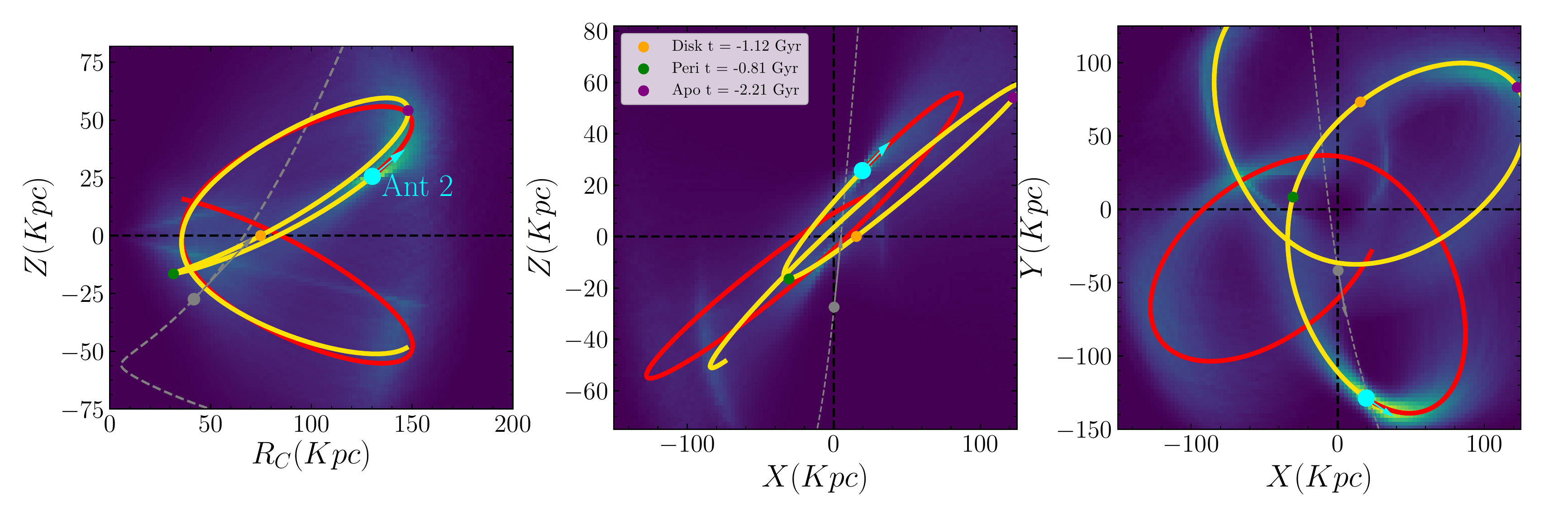}
     \caption{Orbit of Antlia~2 in Galactocentric coordinates. The orbit of
Ant~2 is obtained by integrating for 5 Gyr the initial conditions as recorded
in Table~\ref{tab:Properties} in the {\it MWPotential2014} potential from {\it
galpy} \citep{2015ApJS..216...29B}, but in a DM halo which is 12\% more
massive. The position of Ant~2 is shown in cyan, together with its past
(yellow) and future (red) orbits. The most recent pericentre and apocentre are
marked with green and purple filled circles, respectively. The most recent
``disc'' crossing is shown in orange. Note that this happens about $\sim90$
kpc from the centre of the galaxy. The background density map corresponds to
the cumulative positions of the orbits sampled according to the uncertainties
in the dwarf's line-of-sight velocity, proper motion, and distance. The orbit
and position of the LMC are shown in grey. On this orbit, the pericentre of
Ant~2 is at $37_{-15}^{+20}$ kpc.}
     \label{fig:orbit}
\end{figure*}%

\section{Spectroscopic follow-up}\label{sec:spec}

Immediately after the object's discovery at the Flatiron Gaia Sprint, we
sought to obtain spectroscopic follow up of some of Ant~2's RGB members.  The
night of 24 June 2018 we obtained service mode observations of targets in the
field of Ant~2 with the 2dF+AAOmega Spectrograph~\citep{Sharp2006} on the
3.9~m Anglo-Australian Telescope. The data consist of $3\times 30$ min
exposures taken with an average seeing of 1.4\arcsec, over an airmass range of
1.4 to 2 and at a moon distance of $\sim80\degr$ from the full moon. We used
580V (R $\sim 1300$) and 1700D (R $\sim 10000$) gratings in the blue and red
channels, respectively.  Because of the bright moon conditions during
observations, the signal-to-noise of the blue spectra taken with the 580V
grating was low, and hence we only used the red 1700D spectra -- covering the
wavelength range between 8450\AA\ to 9000\AA\ and containing the infrared
calcium triplet -- for the analysis in this paper. 

The strong unambiguous RGB signal, and the availability of the
colour-magnitude, proper motion and spatial information allowed easy and
efficient target selection. Figure \ref{fig:specres} gives the DECam
colour-magnitude diagram of the targets selected for the spectroscopic
follow-up. Note that the original selection was performed using
\textit{Gaia}'s BP and RP band-passes, which is why the selection deviates
substantially from the isochrone colours at fainter magnitudes. On top of the
CMD-based selection, we also required the targets to have proper motions
within 1.5 mas/yr of $(\mu_\alpha \cos\delta, \mu_\delta) =
(-0.04,-0.04)$mas/year. The targets were selected to occupy the whole of the
2~deg field of view of the 2dF+AAOmega spectrograph. We observed a total of
349 candidate stars in Ant 2.

The data reduction was performed using the latest version of the $2dfdr$
package (v6.46)\footnote{\url{https://www.aao.gov.au/science/software/2dfdr}},
including the following procedures: bias subtraction, 2D scattered light
subtraction, flat-fielding, Gaussian-weighted spectral extraction, wavelength
calibration, sky subtraction, and spectrum combination.



To model the observed spectra and obtain chemical abundances and radial
velocities we use a direct pixel-fitting approach by interpolated spectral
templates \citep[see, e.g.][]{Koleva2009,Koposov2011,Walker2015}. We use the
PHOENIX v2.0 spectral library \citep{Husser2013} that spans a large range of
metallicities (from [Fe/H]$=-4$ to [Fe/H]$=1$, $[\alpha/Fe]$ between $-0.2$
and $1.2$) and stellar atmospheric parameters. For parameter values that fall
between templates, we combine the Radial Basis Function interpolation, which
is used to create a grid with a step size finer than that in the original
PHOENIX grid, with a linear N-d interpolation based on the Delaunay
triangulation \citep{Hormann2014} (at the last stage). At each
spectral-fitting step, the polynomial continuum correction transforming the
template into observed data is determined. As the original template grid has
$\log g$, $\rm{T}_{\rm{eff}}$, $[$Fe/H$]$ and $\alpha$ parameters, we sample
those together with the radial velocity using the emcee
\citep{ForemanMackey2013} ensemble sampler, while assuming uniform priors over
all parameters. The resulting chains for each parameter of interest are then
used to measure various statistics, such as posterior percentiles and standard
deviations, as well as the measures of non-Gaussianity, such as kurtosis and
skewness \citep[as motivated by][]{Walker2015}. The average signal-to-noise
(per pixel) of the spectra is 5.8, and for the spectra with S/N>3 the median
uncertainties were 2.67\,km/s, 0.7 dex, 325 K, and 0.35 dex for the RV, $\log
g$, $\rm{T}_{\rm{eff}}$, and [Fe/H], respectively. Most radial velocity
uncertainties are significantly larger than the systematic floor of 0.5 km/s 
of the 1700D setup on AAOmega (S. Koposov, private communication).
Table~\ref{tab:Properties_sp} records all the relevant information for the
spectroscopic measurements reported here. 

For the most part, the analysis of the stellar kinematics in the paper
utilises the subset of stars with 1$\sigma$ uncertainties in the radial
velocity less than 10 km\,s$^{-1}$ and residual kurtosis and skewness less
than 1 in absolute value in order to ensure that the posterior is close enough
to a Gaussian. The number of these stars is 221. The velocity distribution of
these stars is shown in Figure~\ref{fig:specres}. The distribution reveals a
strikingly prominent peak at $\sim 290$ km\,s$^{-1}$ containing 159 of the 221
stars in the sample - undoubtedly Ant~2's velocity signature - as well as a
broad (and weak) contribution from the MW halo and MW disk.  The association
of the velocity peak with Ant~2 is particularly clear in
Fig.~\ref{fig:rv_feh}, where we show the radial velocities of the observed
stars as a function of their proper motion and spectroscopic metallicity. The
stars in the RV peak have metallicities significantly lower than the field
stars and are concentrated around the proper motion value of $(\mu_\alpha \cos
\delta,\mu_\delta) \approx (0,0)$. In the next section, we model the observed
distribution to measure the kinematic properties of the newly discovered
dwarf.

\subsection{Kinematic modelling}

To describe the kinematics of the system we construct a generative model of
the proper motions and radial velocities. The right two panels of
Figure~\ref{fig:rv_feh} show the data used for the model. We highlight that
both the proper motions and radial velocities are highly informative for
identification of members of Antlia~2, however the proper motion errors are
noticeably larger. For the foreground contaminants, our model assumes a
2-component Gaussian mixture distribution in radial velocity and a uniform
distribution over proper motions within our selection box. The radial velocity
distribution of the Antlia~2 members is modeled by a Gaussian, while the
proper motions are assumed to have no intrinsic scatter and therefore are
described by a delta function as specified below:

\begin{eqnarray}
{\mathcal P}({\bm \mu}, V|\alpha,\delta) = (1-f_{o}) (f_{b,1} N(V | V_{b,1}, \sigma_{b,1}) +
\nonumber \\
 (1-f_{b,1}) N(V | V_{b,2} ,\sigma_{b,2} ) ) U({\bm \mu}) + f_o N(V|V_{o}, \sigma_o)) 
\nonumber \\
  \delta({\bm \mu}-{\bm \mu}_o) 
  \label{eq:rv_model} 
\end{eqnarray}

where $f_o$ is the fraction of stars belonging to Ant~2 and $f_{b,1}$ is the
fraction of the foreground stars belonging to a first Gaussian component,
$V_o$ and $\sigma_o$ are the systemic velocity and the velocity dispersion of
Ant~2, and $V_{b,1}$,$V_{b,2},\sigma_{b,1},\sigma_{b,2}$ are the means and
standard deviations of the Gaussian distribution of the foreground. U(${\bf
\mu})$ is the bivariate uniform distribution within the proper motion
selection region. The Gaussian uncertainties on both proper motions and radial
velocities for each star are easily taken into account in this model by
convolving the distribution with the appropriate Gaussian. The only additional
assumption we make to take into account the uncertainties is that the
contaminants are approximately uniformly distributed over a much larger area
than the proper motion selection area. We note also that the probability
distribution in Eq.~\ref{eq:rv_model} is conditioned on Right Ascension and
Declination, as some of the variants of the model described below consider the
dependence of the systemic velocity and the proper motion $V_o$ and $\mu_o$ on
the star's position.

Because Ant~2 is exceptionally extended on the sky, we consider a situation
where the systemic velocity of the object can spatially vary across the
object. Such velocity field evolution could be induced either by the internal
dynamics in Ant~2 or the perspective `rotation' effect due to the proper
motion of the object \citep{Merritt1997,Kaplinghat2008,Walker2008}. To test
these possible scenarios above we consider the following three models for the
systemic velocity $V_o$ of Ant~2:

\begin{itemize}
\item Constant radial velocity ($V_o$)
\item Radial velocity is a function of systemic velocity, proper
  motion and position of the star $V_o = V_o(V_{o,0}, \alpha, \delta,
  D, \mu_{\alpha,0}, \mu_{\delta,0})$ as predicted by projection
  effects (perspective rotation).
\item The previous model combined with the linear gradient in radial
  velocity $V_o = V_{o,0} + V_x (\alpha-\alpha_0) \cos \delta_0 + V_y
  (\delta - \delta_0) $
\end{itemize}

We also ran the second model using the radial velocity data only, while
ignoring the \textit{Gaia} proper motion information in order to separate the
inference of Ant~2's proper motion driven by the \textit{Gaia} data from the
inference based on the radial velocity gradients.

\begin{figure}
    \includegraphics[width=0.5\textwidth]{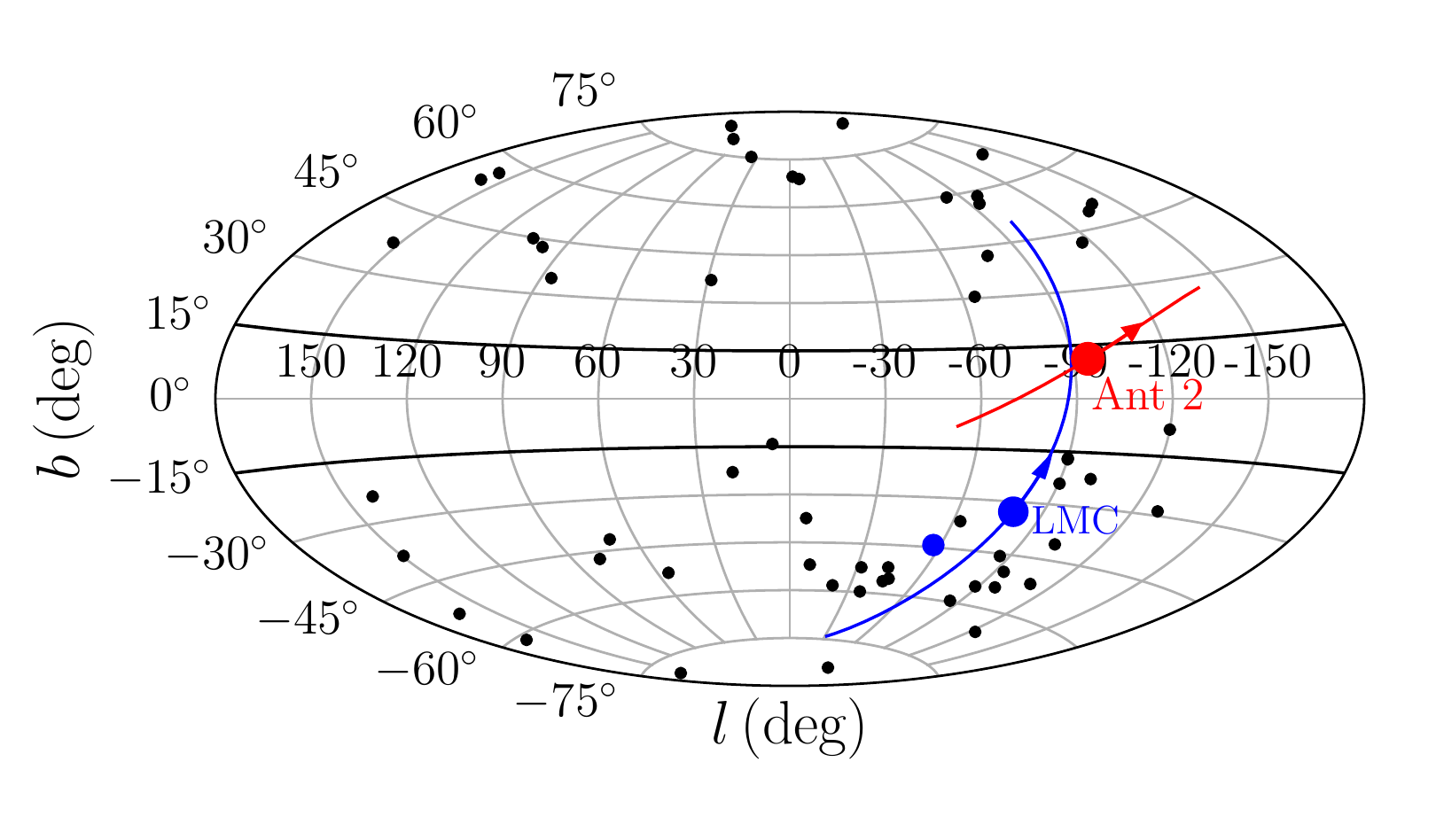}
    \caption{Distribution of the MW dwarf satellites in Galactocentric
      coordinates (from \citet{McConnachie2012} plus Crater~2
      \citep{2016MNRAS.459.2370T}, Aquarius~2
      \citep{2016MNRAS.463..712T}, DESJ0225+0304
      \citep{2017MNRAS.468...97L}, Pictor~II
      \citep{2016ApJ...833L...5D}, Virgo~I
      \citep{2016ApJ...832...21H}, Cetus~III
      \citep{2018PASJ...70S..18H}, Car~II and Car~III
      \citep{2018MNRAS.475.5085T}, Hydrus~1
      \citep{2018MNRAS.tmp.1695K}, and updated values for the Andromeda galaxies from \citet{2016ApJ...833..167M}). The position of Ant~2 is shown as a red filled
      circle. The positions of the Magellanic clouds are shown in
      blue. Other MW dwarf galaxies are shown in black. The red and
      blue lines are the orbits of Ant~2 and the LMC,
      respectively. Black lines enclose the Galactic plane between
      $b\pm15$, highlighting the ZOA, which presently clearly shows a
      dearth of MW satellites (Note that the black dot closer to the disk corresponds to Canis Major, whose classification as galaxy is uncertain). Interestingly, while the LMC's orbit is
      not aligned with that of Ant~2, the new object lies close to the
      projection of the Cloud's orbital path. Note however that
      testing the possibility of association between these two objects
      is not feasible without a detailed simulation of the Magellanic Clouds and
      Ant~2's accretion onto the MW (see also Fig.\ref{fig:orbit}).}
    \label{fig:fullsky}
\end{figure}

For all of the three models described above the parameters were sampled using
the ensemble sampler.  For each posterior sample we ignored the first half of
the chain as a burn-in/warm-up. For the remainder of the chains we verified
the convergence by checking the acceptance rate across walkers and verified
that the means and the standard deviations of the first third of the chains
agreed well with the last third part of the chains for each parameter
\citep{Geweke1991}. The values of common parameters measured from different
models mostly agree within 1$\sigma$.  The parameter values from the model
with perspective rotation and no intrinsic velocity gradient such as systemic
velocity, proper motions, and velocity dispersion are given in the bottom part
of Table~\ref{tab:Properties}. The main results are the following: the
systemic velocity is $V_o =290.7\pm 0.5$\, km/s, with a velocity dispersion of
$\sigma_o = 5.71\pm1.08$\,km/s and a systemic proper motion of
$\mu_{\alpha}\cos\delta= -0.095\pm0.018$ mas/yr, $\mu_{\delta} =
0.059\pm0.024$ mas/yr. Note that additional systematic uncertainties of
$0.030$ and $0.036$ for $\mu_{\alpha}$ and $\mu_{\delta}$ could be considered,
but Ant~2 is not likely to be affected by these systematics as, given its
large angular extent, they should average out \citep{2018A&A...616A..12G}.
Also note that we assume a zero binary fraction when estimating the velocity
dispersion, which could have the effect of biasing its measurement to slightly
higher values \citep[see, e.g.,][]{2017AJ....153..254S}.

\begin{figure}
     \includegraphics[width=0.5\textwidth]{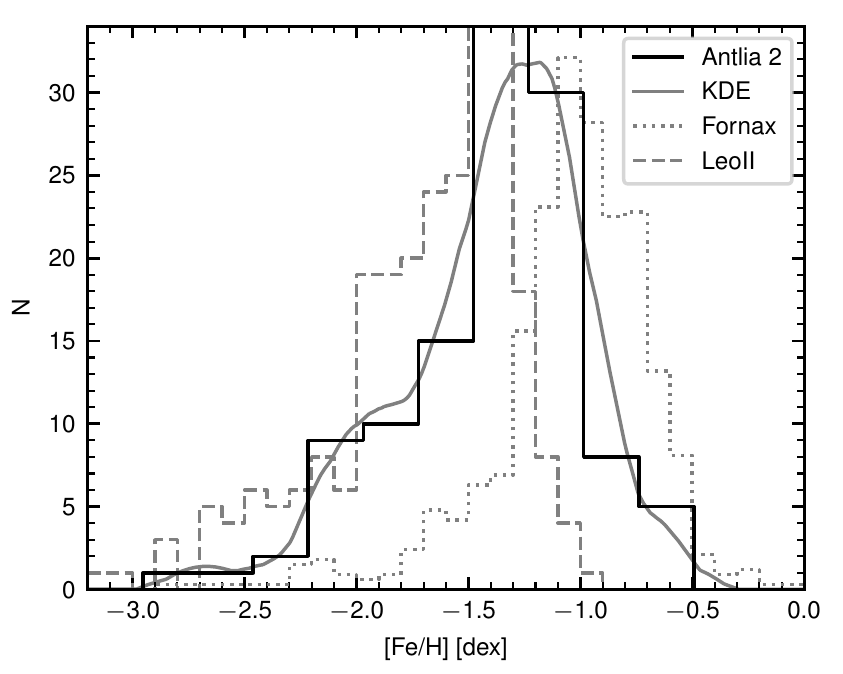}
     \caption{Distribution of spectroscopic metallicities of Ant~2's
       stars with 274$<$HRV$<$303\,km/s and metallicity error less
       than 0.5 dex. The black line is the [Fe/H] histogram with a bin
       size of 0.2 dex, while the grey curve shows the kernel density
       estimation using the Epanechnikov kernel. The mean metallicity
       is $<[Fe/H]>$=-1.36, with a significant scatter toward low
       metallicities. Metallicity distributions of stars in Fornax and
       Leo II are also shown for comparison.}
     \label{fig:fehhist}
\end{figure}

The goodness-of-fit (log-likelihood) values for the different models listed
above were comparable, with a likelihood ratio of $\sim 1$ - indicating that
no very strong evidence for perspective rotation or intrinsic rotation was
observed. However the model that was applied to the radial velocity data, while
ignoring the \textit{Gaia} proper motions, implied a kinematic proper motion of
$\mu_\alpha \cos \delta,\mu_\delta=-0.26\pm0.13,-0.28+/-0.10$ mas/yr, which is
in some tension with the overall (Gaia-based) proper motion of the system.
Figure~\ref{fig:pms} shows the comparison between the inferred systemic proper
motion values, as well as the expected proper motion direction if it was
aligned with the orientation of the Antlia~2's iso-density contours. The most
likely explanation for the mismatch of the kinematic proper motion and the
astrometric proper motion is that Antlia has some intrinsic velocity gradient.
This can be associated either with the tidal disruption of the system or with
intrinsic rotation.

We have also attempted to measure the velocity dispersion gradient in Antlia~2
by applying the model in Eq.~\ref{eq:rv_model} to stars in 3 different angular
distance bins (with respect to Antlia~2's center). The bins were selected such
that they have an approximately equal number of stars. We kept the parameters
of the foreground velocity and the proper motion distribution fixed across
those bins and only allowed the velocity dispersion of the dwarf and a mixing
fraction of dwarf stars to change from bin to bin. The results of this model
are shown in Fig.~\ref{fig:veldisp_grad}.  We can see that the velocity
dispersion in the very central bin is measured to be somewhat lower than in
the outer bins, although by only $\sim$ 2 $\sigma$. While the stellar velocity
dispersion in a dark matter dominated system can change with radius, there are
other possible explanations. Apart from a random fluctuation, this could be
due to the existence of a velocity gradient associated with either rotation or
tidal disruption, which would tend to inflate the outer velocity dispersion
measurements.
        

\subsection{Orbit}\label{sec:orbit}

\begin{figure*}
    \includegraphics[width=\textwidth]{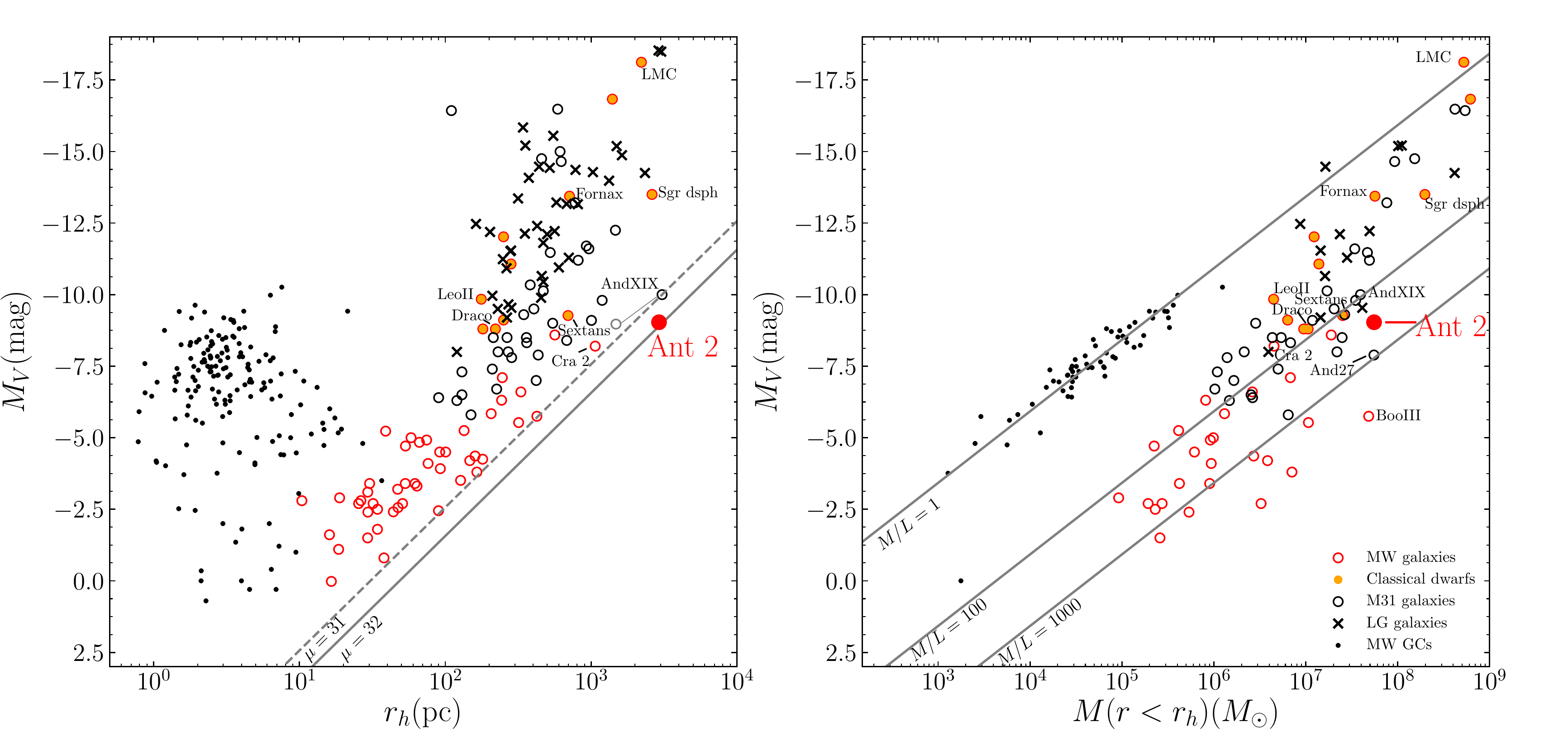}
    \caption{{\it Left:} Absolute magnitude as a function of
      half-light radius for galaxies in the Local Group as well as the
      MW globular clusters. Orange filled circles show the positions
      of the Classical dwarfs, open red circles give the positions of
      fainter MW satellites (see Figure~\ref{fig:fullsky} for references to the MW satellite values used) and black dots show the positions of
      the MW GCs
      \citep{Harris2010,Belokurov2010,Munoz2012,Balbinot2013,Kim2015a,Kim2015b,Kim2016,Laevens2015,Weisz2016,2016MNRAS.458..603L,2017MNRAS.468...97L,2016ApJ...830L..10M,2017MNRAS.470.2702K,2018MNRAS.478.2006L};
      the M31 satellites are shown as black open circles and the other
      LG galaxies are shown as black crosses \citep[both
        from][]{McConnachie2012}. The position of Antlia~2 is shown
      with a red filled circle. With a size similar to the LMC, but
      the luminosity close to that of the faintest of the classical
      dwarfs, Ant~2 has a surface brightness more than 1 magnitude
      fainter than any previously known galaxy. {\it Right:} Object
      luminosity as a function of the dynamical mass within the
      half-light radius. Symbols are the same as in the right panel,
      but only systems with known velocity dispersions are
      shown. Masses were estimated using the relation from
      \citet{Walker2009}. The solid grey lines correspond to
      mass-to-light ratios of 1, 100, and 1000.}
    \label{fig:mvrh}
\end{figure*}

We apply the kinematics of Ant~2 obtained above to gauge the satellite's
orbital properties using galpy\citep{2015ApJS..216...29B}. Motions are
converted to the Galactic standard of rest by correcting for the Solar
rotation and the local standard of rest velocity, using $v_{circ}=220$\,km/s
and $v_{lsr}=(11.1,12.24, 7.25)$\,km/s \citep{uvw}. Figure~\ref{fig:orbit}
presents the orbit of Ant~2 generated using the {\it MWPotential2014}
\citep{2015ApJS..216...29B} with the MW's halo mass increased from
$0.8\times10^{12}\,M_\odot$ to $0.9\times10^{12}\,M_\odot$ \citep[see,
e.g.][for recent mass estimates]{2018arXiv180709775V}. The density map in the
figure shows the accumulation of the orbits with initial conditions sampled
using the uncertainties in radial velocity, proper motions, and distance. The
current position of Ant~2 is shown in cyan, and its past and future orbits are
shown in black and yellow, respectively. We then estimate the pericenter --
using all the sampled orbits -- to be at $37_{-15}^{+20}$ kpc, which is just
close enough to induce some tidal disruption in the satellite (see the
discussion of the mass measurement of Ant~2 in Section~\ref{sec:conc}). A
higher MW mass of $1.8\times10^{12}$ \citep[e.g.][]{2018arXiv180411348W}
reduces the median pericenter to $\sim 25_{-9}^{+13}$\,kpc, which is around
the lower limit of the pericenter uncertainties found for the orbits in a
lighter MW \citep[see, however,][ for a possible bias toward larger pericenter
values using this method]{2008ApJ...678..187V}. According to the orbit
computed, Ant~2 last passed through pericentre about 1 Gyr ago.  It recently
crossed the plane of the Galactic disc, but 95 kpc away from the MW centre.
The dwarf is about to reach its apocentre. For comparison, the Figure also
shows the orbit of the LMC (in grey). Both galaxies have similar directions of
motion; in fact, Ant~2 is currently sitting very close to where the LMC is
heading and where \citet{Jethwa2016} predict a large number of low-mass dwarfs
stripped from the LMC. However, given the significantly slower orbital
velocity of Ant~2 - resulting in a significantly different orbital phase - any
obvious association between the two objects seems somewhat unlikely.
Figure~\ref{fig:fullsky} illustrates the difference in the orbital paths of
the LMC and Ant~2 more clearly.

\subsection{Chemistry}

\begin{figure*}
    \includegraphics[width=\textwidth]{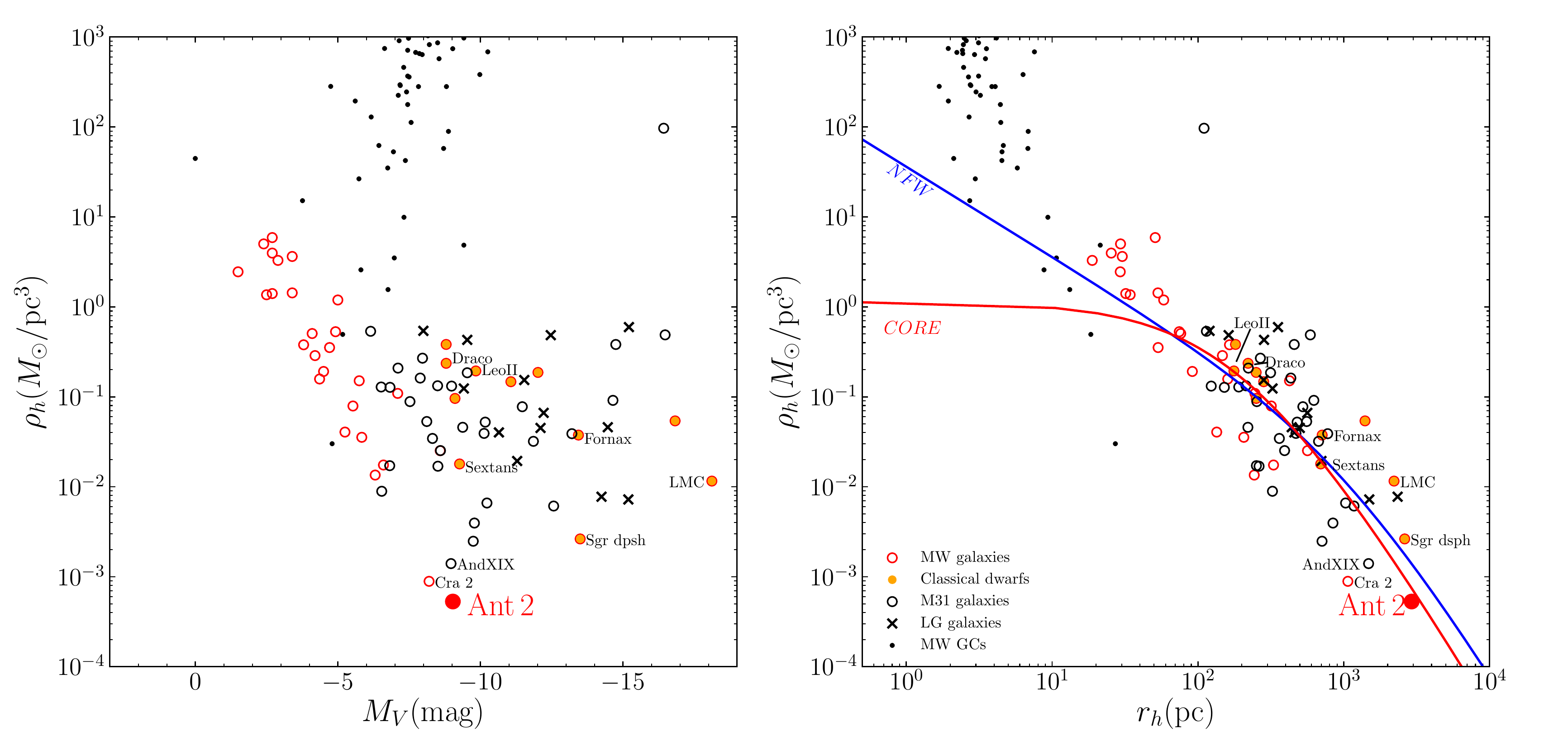}
    \caption{Average (or effective) total matter density within the
      half-light radius for stellar systems in the LG. Symbols are the
      same as in Figure~\ref{fig:mvrh}. {\it Left:} Average density as
      a function of the object's absolute magnitude. A clear selection bias can be seen in the lower left
      corner of the plot, in which galaxies become too diffuse to be
      detected. {\it Right:} Effective density as a function of
      half-light radius. The blue and red lines are the best-fit
      ``universal profiles'' for dwarf galaxies from
      \citet{Walker2009}. Blue is an NFW profile, and red corresponds
      to a cored double power law model. Ant~2 extends the trends
      reported in \citet{Walker2009} to larger sizes.}
    \label{fig:rhden}
\end{figure*}

We also look at the metallicity distribution of likely members of Ant~2. We
select stars within 15 km\,s$^{-1}$ of the systemic velocity of Ant~2, small
radial velocity error $\sigma_v<5$ and small uncertainty on [Fe/H]$<0.5$ and
residual kurtosis and skewness on [Fe/H] less than 1. The stellar metallicity
distribution of this sample - which we believe to be free of contamination -
is shown in Figure~\ref{fig:fehhist}. We note that Ant~2's metallicity peaks
at $[{\rm Fe/H}]=-1.4$, i.e.  noticeably higher than the majority of
ultra-faint dwarf galaxies. Additionally, the [Fe/H] histogram also shows some
moderate asymmetry towards low metallicities, which has been seen in other
objects \citep{Kirby2010}. Accordingly, for comparison we over-plot the
metallicity distributions of a couple of classical dwarf spheroidal galaxies (
Leo II and Fornax ) from \citet{Kirby2010}, whose overall metallicities and MDF
shapes are not dissimilar to those of Ant~2. Leo II also has a stellar mass similar to Ant~2's.

To measure the mean metallicity of the system we fit the metallicity
distribution by a Gaussian mixture with two Gaussians (due to possible
asymmetry of the MDF). In the modeling we take into account abundance
uncertainties of individual stars. The resulting mean metallicity is
[Fe/H]$=-1.36\pm 0.04$, with a standard deviation of $\sigma_{\rm
[Fe/H]}=0.57\pm0.03$. These measurements are provided in the
Table~\ref{tab:Properties}. We also note there is a possible abundance
gradient with radius, as the subset of stars within 0.5 degrees has a mean
metallicity of ${\rm [Fe/H]}=-1.29\pm 0.05$ and the stars outside $0.5$
degrees have a mean metallicity of ${\rm [Fe/H]=-1.44\pm0.06}$. The difference
in metallicities is only marginal, however it is not unexpected, as similar
trends with centrally concentrated more metal-rich stellar populations have
been observed in many (especially classical) dwarf galaxies
\citep{Harbeck2001,Koch06}.

\section{Discussion}\label{sec:discuss}

Figure~\ref{fig:mvrh} presents the physical properties of Ant~2 in comparison
to other stellar systems in the MW and the Local Group. The left panel shows
the distribution of intrinsic luminosities (in absolute magnitudes) as a
function of the half-light radius. Strikingly, no other object discovered to
date is as diffuse as Ant~2. For example, the so-called Ultra Diffuse Galaxies
\citep[][]{UDG} have sizes similar to Ant~2, but are typically $\sim6$ mag
brighter. Overall, compared to systems of similar luminosity, the new dwarf is
several times larger, while for objects of comparable size, it is $\sim3$
orders of magnitude fainter. One exception to this is And~XIX
\citep{Mcconnachie2008}, which has had its size updated in
\citet{2016ApJ...833..167M} from 6.2\arcmin to 14.2\arcmin. In the figure, we
show the old measurement connected with a line to the new measurement. As we
can see, And~XIX is similar in size to Ant~2, but $\sim 2$ magnitudes
brighter. While the half-light radius of And XIX has recently been updated to
a much larger value, the available spectroscopy only probes the mass
distribution within a much smaller aperture, corresponding to the earlier size
measurement. We therefore report a mass measurement for And XIX within
6.2\arcmin but caution the reader against over-interpreting this number. The
right panel of the Figure shows the satellite's luminosity $M_V$ as a function
of the mass within the half-light radius for systems with known velocity
dispersion. Here, we have used $M_{dyn}=581.1\,\sigma_rv^2\,r_h$, the mass
estimator suggested by \citet{Walker2009}.  Superficially, Ant~2, although
sitting at the edge of the distribution of the currently known dwarfs, does
not appear as extreme in the plane of absolute magnitude and mass within the
projected half-light radius.

However, the similarity of the Ant~2's mass to that of other dwarfs of
comparable luminosity is clearly deceptive. This is because the dwarf's
half-light radius is typically an order of magnitude larger than that of other
objects at the same level of brightness. Figure~\ref{fig:rhden} illustrates
this by showing the effective density of each satellite, in other words the
mass within the half-light radius divided by the corresponding volume. The
left panel gives the density as a function of the intrinsic luminosity, while
the right panel displays density as a function of the half-light radius. As
the Figure convincingly demonstrates, Ant~2 occupies the sparsest DM halo
detected to date. Interestingly, the dwarf appears to extend the ``universal''
density profile suggested by \citet{Walker2009} to lower densities. At the
radius probed by Ant~2, the cuspy (blue line) and cored (red line) density
profiles start to decouple appreciably. The satellite seems to follow the red
curve within the observed scatter. Could the extremely low stellar and DM
densities in Ant~2 be the result of the tidal influence of the MW? While the
satellite does not come very close to the Galactic centre (as discussed in
Section~\ref{sec:orbit}), at its nominal pericentre of $\sim$40 kpc, the MW's
density is around twice the effective (half-light) density of the object
\citep[assuming the Galactic mass measurement
of][]{Williams2017}\footnotemark, hence some amount of tidal
heating/disruption would be expected. Bear in mind however that, while tides tend to
lower a satellite's density, typically \citep[as demonstrated
by][]{Penarrubia2008, Penarrubia2012}, as the satellite loses mass to the
host, it tends to shrink rather than expand. This would imply that the dwarf
started with an even {\em larger} half-light radius \citep[also
see][]{Sanders2018}.

\footnotetext{ At 20 kpc, the lower $1\sigma$ limit of Ant~2's pericenter, the
MW density goes up to $\sim5$ times Ant~2's effective density.}

\begin{figure}
     \includegraphics[width=0.5\textwidth]{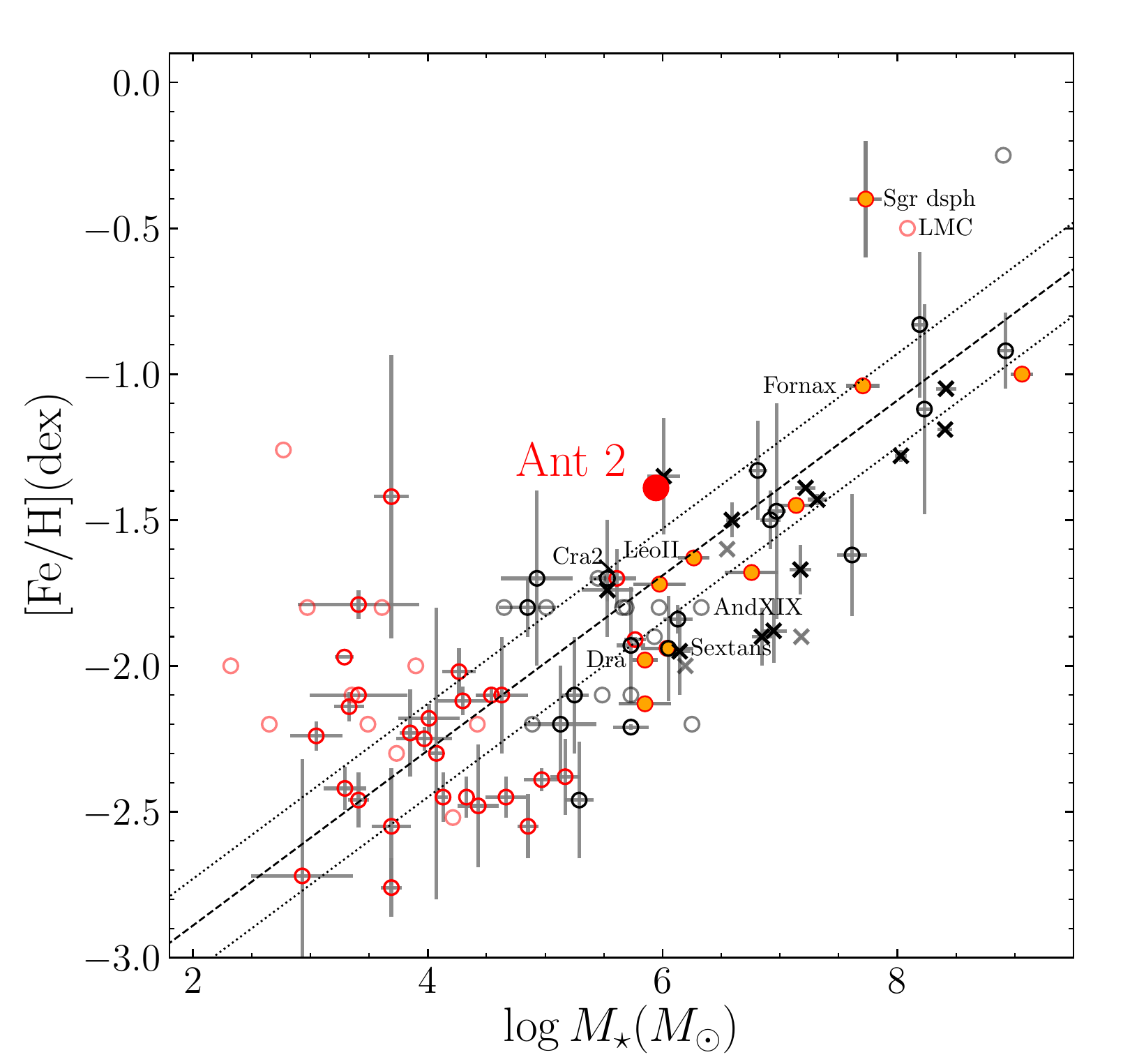}
    \caption{Metallicity as function of stellar mass. The symbols
      are the same as in Figure \ref{fig:mvrh}. Objects with
      spectroscopic metallicities are in bold, while those with only
      photometric measurements are shown as faded symbols. The dashed black line is
      the mass-metallicty relation from \citet{2013ApJ...779..102K}
      with the corresponding 1$\sigma$ scatter indicated by the dotted
      lines. Spectroscopic metallicities for dwarf galaxies are from
      \citet{McConnachie2012,2013ApJ...779..102K,2015ApJ...810...56K,2017ApJ...838...83K,2017ApJ...839...20C,2016MNRAS.463..712T,2017MNRAS.467..573C,2018MNRAS.tmp.1695K,2015ApJ...811...62K,2018ApJ...857..145L,2017ApJ...838....8L,2017ApJ...834....9K,2016ApJ...819...53W,2017ApJ...838...11S,2016ApJ...833...16K}. The
      remaining photometric data are from the same sources as in
      Figure~\ref{fig:mvrh}.}\label{fig:massmet}
\end{figure}

There are other indirect hints that Ant~2 might not have escaped the MW's
tides. The RR Lyrae used to identify the object in \textit{Gaia} DR2 data lie
significantly closer to the observer than the dwarf itself, as traced by the
RGB and BHB populations (see Section~\ref{s:bhb} for details). The nominal
mean distance to these stars is $\sim80$ kpc, implying that they are some 50
kpc away from the dwarf, signaling extended tidal tails. Moreover, as
Figure~\ref{fig:pms} illustrates, another clue is the close alignment between
the direction of the dwarf's motion (as measured using the GDR2 data, black
filled circle with error-bars) and the elongation of its iso-density contours
(dashed line). In addition, the radial velocity gradient (grey filled circle
with error-bars) does not fully match the \textit{Gaia} proper motions,
possibly indicating a velocity field affected by rotation or tides. One can
also look for signs of disruption in the mass-metallicity diagram. Figure
\ref{fig:massmet} shows metallicity as a function of stellar mass for objects
in the Local Group. Systems with metallicity inferred from spectroscopy are
plotted in full colour, while metallicities deduced from photometry alone are
shown as light grey symbols. We have also completely removed objects with
metallicities drawn solely based on the colour of the RGB
\citep{1990AJ....100..162D}, as these are more affected by systematics due to
the age/metallicity degeneracy compared to other photometric methods
\citep[see e.g.][for further discussion]{Mcconnachie2008,2013ApJ...779..102K}.
Assuming that a correlation exists between the galaxy's metallicity and its
stellar mass - as Figure~\ref{fig:massmet} appears to indicate - objects that
have suffered any appreciable amount of mass loss would move off the main
sequence to the left in this plot. Ant~2 is indeed one such example: while not
totally off the mass-metallicity relation, it clearly hovers above it, thus
suggesting that some tidal disruption might have occurred. Indeed, if one
assumes Ant~2's metallicity originally fit this relation and its metallicity
has remained constant, one would expect that Ant~2 should have initially had a
stellar mass of $(1\pm0.4)\times10^7 M_\odot$, which would imply that today we
are seeing only $\sim9\%$ of its original population.

\subsection{Dark Matter Halo}\label{sec:halo}

Since the Gaia DR2 data do not resolve the internal proper motion
distributions within Ant 2, inferences about dynamical mass must rely on the
projection of phase space that is sampled by star counts on the sky and
spectroscopic line-of-sight velocities.  These observations are usefully
summarized by the global velocity dispersion, $\sigma_{\rm rv}=5.7\pm 1.1$ km
s$^{-1}$, and halflight radius, $R_{\rm h}=2.86\pm 0.31$ kpc.  On dimensional
grounds, the dynamical mass enclosed within a sphere of radius $r=\lambda
R_{\rm h}$ can be written

\begin{equation}
M(\lambda R_{\rm h})=\frac{\lambda \mu}{G} R_{\rm half}\,\sigma^2_{\rm rv}.
\label{eq:estimator}
\end{equation}

Equating $\sigma_{\rm rv}$ with the global mean (weighted by surface
brightness) velocity dispersion, the coefficient $\mu$ depends only on the
gravitational potential and the configuration of tracer particles, via the
projected virial theorem \citep{agnello12,errani18}:

\begin{equation}
\sigma^2=\frac{4\pi G}{3}\displaystyle\int_0^{\infty}r\nu(r)M(r)dr,
\label{eq:projvirial}
\end{equation}

where $\nu(r)$ is the deprojection of the projected stellar density profile;
for the adopted Plummer profile, $\nu(r)\propto (1+r^2/R_{\rm h}^2)^{-5/2}$.  

Without invoking a specific mass profile, the simple mass estimator of
\citet{Walker2009} effectively assumes $\lambda=1$ and $\mu=5/2$, implying for
Ant 2 a dynamical mass $M(R_{\rm h})\approx [5.4\pm 2.1]\times 10^7 M_{\odot}$
enclosed within a sphere of radius $r=2.9$ kpc; the quoted uncertainty
reflects only the propagation of observational errors, neglecting systematic
errors that recent simulations suggest tend to be $\la 20\%$ regardless of
stellar mass \citep{campbell17,Gonzalez17}.  The more recent estimator of
\citet{errani18}, calibrated to minimize systematics due to uncertainty about
the form of the mass profile, uses $\lambda=1.8$ and $\mu=3.5$, implying a
dynamical mass $M(1.8R_{\rm h})\approx [1.37\pm 0.54]\times 10^8 M_{\odot}$
enclosed within a sphere of radius $r=5.2$ kpc. The corresponding estimator of
\citet{campbell17} has $\lambda=1.91$ and $\mu=3.64$, giving a mass $[1.53\pm
0.61]\times 10^8 M_{\odot}$ enclosed within a sphere of radius $r=5.6$ kpc.

In contrast to the use of mass estimators, specification of the stellar number
density ($\nu(r)$) and enclosed mass ($M(r)$) profiles lets one use Equation
\ref{eq:projvirial} to calculate the global velocity dispersion
\textit{exactly}. Even though Ant~2's profile might have been modified by its
interaction with MW, in order to place Ant 2 in a cosmological context, we
first consider the properties of dark matter halos that might host Ant 2 while
following the NFW enclosed-mass profile that characterizes halos formed in
N-body simulations \citep{NFW1996,navarro97}:

\begin{equation}
  M_{\rm NFW}(r)=4\pi r_s^3\rho_s\biggl [\ln\biggl(1+\frac{r}{r_s}\biggr ) -\frac{r/r_s}{1+r/r_s}\biggr ].
  \label{eq:nfw}
\end{equation}

An NFW halo is uniquely specified by parameters $M_{200}\equiv M(r_{200})$,
the mass enclosed within radius $r_{200}$, inside which the mean density is
$\langle \rho\rangle_{200}\equiv 200[3H_0^2/(8\pi G)]$, and concentration
$c_{200}\equiv r_{200}/r_s$.  

For Ant 2 and each of the other Local Group dSphs with measured velocity
dispersions and half-light radii (assumed to correspond to Plummer profiles),
we use Equation \ref{eq:projvirial} to find the parameters of NFW halos that
exactly predict the observed velocity dispersion.  For each dwarf, the two
degrees of freedom in the NFW profile result in a `degeneracy curve' of
$M_{200}$ as a function of $c_{200}$ \citep{penarrubia08}; in general, higher
concentrations require lower halo masses in order to predict the same global
velocity dispersion.  

\begin{figure*}
  \centering  
  \begin{tabular}{@{}cccc@{}}
     \includegraphics[width=2.3in, trim=0.2in 1.4in 2.75in 0.4in,clip]{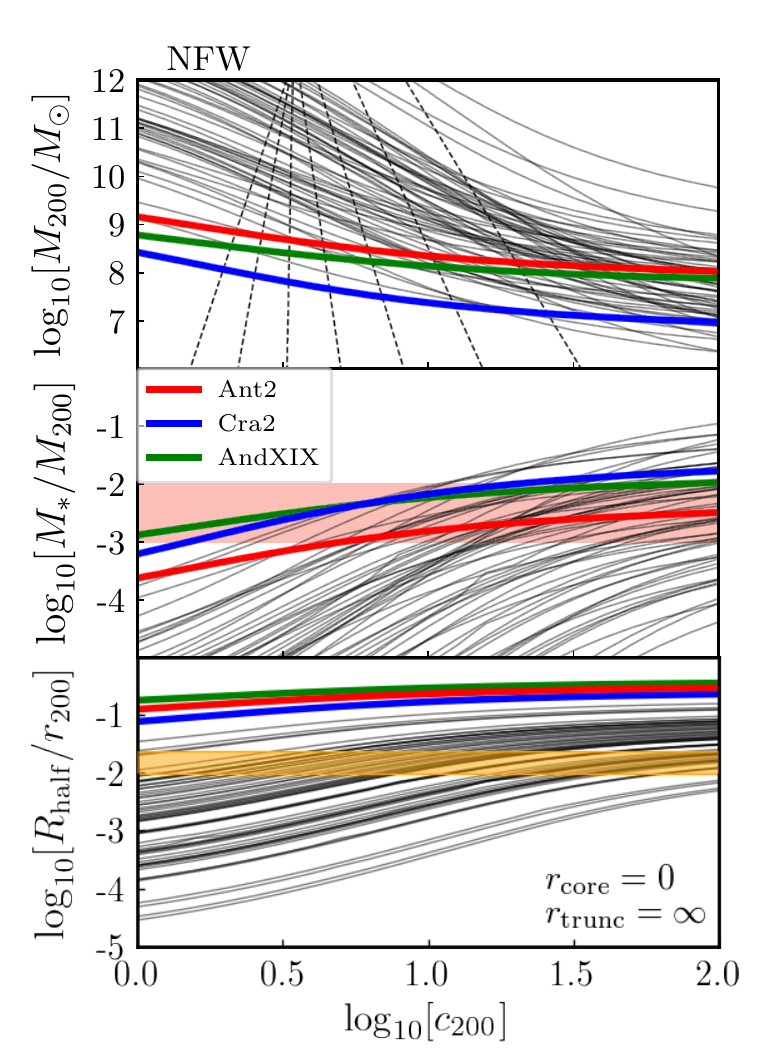}&\includegraphics[width=2in, trim=0.6in 1.4in 2.75in 0.4in,clip]{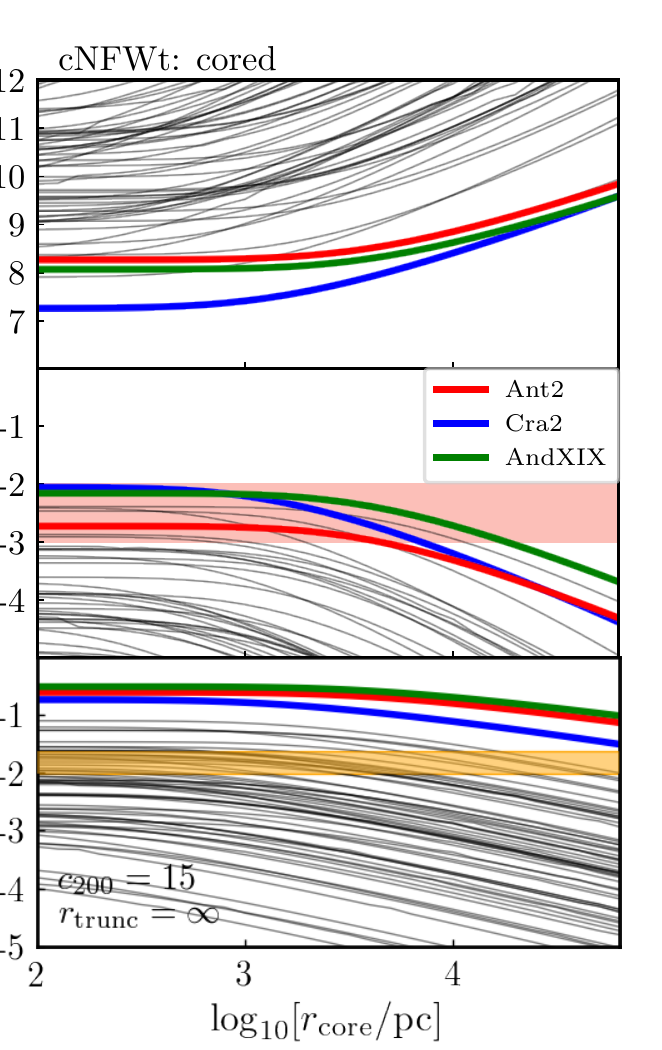}&\includegraphics[width=2in,trim=0.6in 1.4in 2.75in 0.4in,clip]{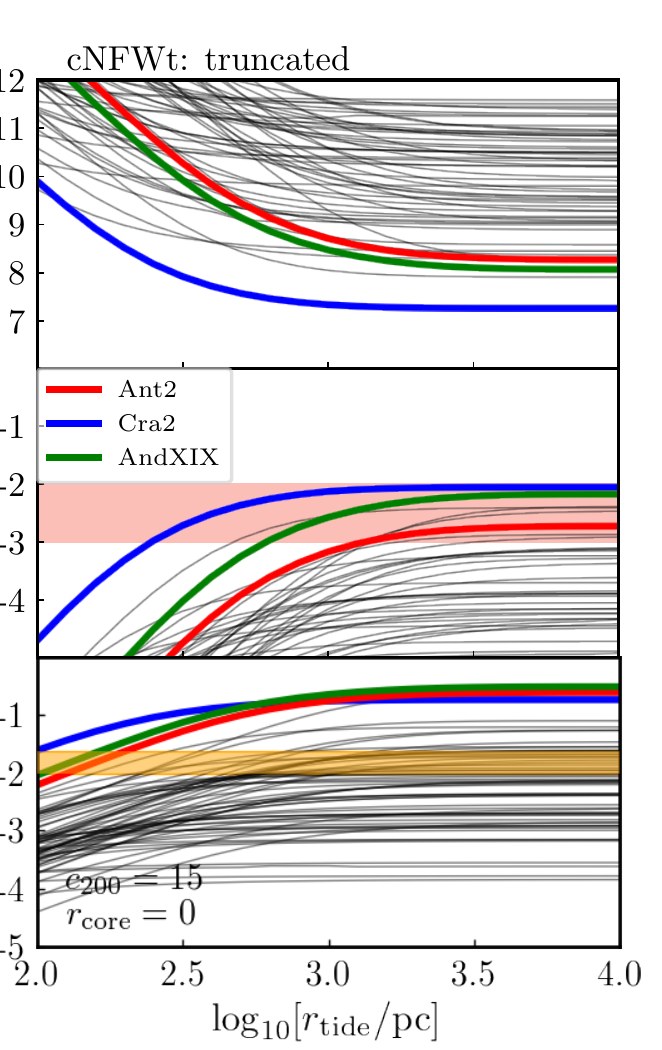}
   \end{tabular}
  \caption{Degeneracy curves showing the relation between dark matter halo
properties and halo mass (top), stellar-to-halo mass ratio (middle) and
halflight radius to halo radius (bottom), for dark matter halos constrained by
the observed half-light radii and stellar velocity dispersions of individual
dwarf spheroidals (one curve per observed dwarf).  In the left panels, the
halos are assumed to exactly follow the NFW form, varying only with halo
concentration $c_{200}$.  In the middle and right panels, the halo profile is
generalised, using Equation \ref{eq:corenfwtide} to allow for modification of
an NFW progenitor by the formation of a central core of radius $r_{\rm core}$
or by tidal truncation beyond radius $r_{\rm trunc}$, respectively (in both
cases, the halo concentration is held fixed at $c_{200}=15$).  In the top-left
panel, dashed lines indicate the halo mass-concentration relations that
describe field halos at redshifts $z=0,1,\ldots,6$ (right to left), derived
from the cosmological N-body simulations of \citet{dutton14}. In the middle
panels, the salmon shaded region indicates the range of mass ratios where
feedback from star formation is expected to transform NFW cusps into cores. In
the bottom panels, the orange shaded region represents the relation
$\log_{10}[R_{\rm h}/r_{200}]=-1.8 \pm 0.2$, which is expected if half-light
radius is determined by halo angular momentum \citep{kravtsov13}.  }
  \label{fig:massconcentration}
\end{figure*}

The top-left panel of Figure \ref{fig:massconcentration} shows the NFW
degeneracy curve for each dwarf galaxy.  Given the measured luminosities, and
assuming a stellar mass-to-light ratio $\Upsilon_*=2M_{\odot}/L_{V,\odot}$,
the middle-left panel shows the corresponding relationship between
concentration and the ratio of stellar to halo mass, $M_{*}/M_{200}$.  The
bottom-left panel shows the relationships between concentration and the ratio
of half-light radius to the halo radius $r_{200}$.

We find that Ant 2 joins Crater 2 \citep{2016MNRAS.459.2370T} and
Andromeda XIX \citep{Mcconnachie2008} as extreme objects amongst the
Local Group dwarfs.  All three are relatively large ($R_{\rm h} \ga 1$
kpc) and cold ($\sigma_{\rm rv}\la 5$ km s$^{-1}$).  As a result,
their plausible NFW host halos tend to have low mass ($M_{200}\la 10^9
M_{\odot}$) even at low concentration.  Perhaps most strikingly, all
three have extremely large ratios of half-light to halo radius, with
$\log_{10}[R_{\rm h}/r_{200}]\ga -1$, putting them $\ga 4\sigma$ above
the the average relation $\log_{10}[R_{\rm h}/r_{200}]=-1.8
\pm 0.2$ describing sizes of the entire galaxy population in the abundance matching scheme of 
\citet{kravtsov13}, suggesting that such a relation might not hold for
  extreme cases like Ant~2.

Larger halo masses and thus smaller ratios of $R_{\rm h}/r_{200}$ in any of
Ant2, Cra2 and AndXIX would require non-NFW halos.  In general there are two
different ways that NFW halo progenitors with more `normal' values of
$M_{200}\ga 10^9M_{\odot}$ and $\log_{10}[R_{\rm h}/r_{200}]\la -1$ might have
been transformed by astrophysical processes into non-NFW halos that would
accommodate the large sizes and small velocity dispersions observed for these
galaxies today.  The first is the outward migration of central dark matter in
response to the rapid loss of gas mass following supernova explosions
\citep[e.g.,][]{Navarro1996,Pontzen2012}.  Recent hydrodynamical simulations
demonstrate that such feedback from galaxy formation can turn primordial NFW
`cusps' into `cores' of near-uniform dark matter density
\citep[e.g.][]{zolotov12,madau14,read16}.    

The second mechanism is mass loss due to tidal stripping, as all three of the
extreme objects are satellites of either the Milky Way or M31.  Indeed, the
orbit inferred from Gaia DR2 proper motions of Crater 2 is consistent with
$r_{\rm peri}\la 10$ kpc \citep[see, e.g.,][]{2018A&A...619A.103F}, compatible
with significant mass loss.  Moreover, \citet{collins14} speculate that tidal
stripping is the cause of AndXIX's extreme kinematics.  In addition to the
loss of both dark and (eventually) stellar mass, consequences of tidal
stripping include steepening of the outer density profile, shrinking of the
luminous scale radius, and reduction of the internal velocity dispersion
\citep{Penarrubia2008,errani17}.  

In order to investigate both of these mechanisms, we consider the generalized
`coreNFWtides' (cNFWt) dark matter halo density model formulated by
\citet{read18}, in which the enclosed mass profile is modified from the NFW
form according to the following:

\begin{equation}
M_{\rm cNFWt}(r) =
\left\{
\begin{array}{ll}
M_{\rm NFW}(r)f^{n_{\rm core}} & r < r_t  \vspace{4mm}\\
M_{\rm NFW}(r_t)f^{n_{\rm core}} \,\, + & \\
4\pi \rho_{\rm cNFW}(r_t) \frac{r_t^3}{3-\delta}
\left[\left(\frac{r}{r_t}\right)^{3-\delta}-1\right] & r > r_t
\end{array}
\right.
\label{eq:corenfwtide}
\end{equation} 

Here, $f^{n_{\rm core}}\equiv \bigl [ \tanh\bigl (r/r_{\rm core}\bigr
)\bigr]^{n_{\rm core}}$ flattens the density profile at radii $r<r_{\rm
core}$, generating a more uniform-density core as $n_{\rm core}$ increases
from $0$ to $1$.  Beyond the truncation radius $r_t$, the density profile
steepens from the NFW outer slope of $\rho(r\gg r_s)\propto r^{-3}$ to
$\rho(r\gg r_t)\propto r^{-\delta}$, with $\rho_{\rm cNFW}(r)=f^{n_{\rm
core}}\rho_{\rm NFW}(r)+\frac{n^{\rm core}f^{n-1}(1-f^2)}{4\pi r^2r_{\rm
core}}M_{\rm NFW}(r)$ and $\rho_{\rm NFW}(r)=(4\pi r^2)^{-1}dM_{\rm NFW}/dr$.

For each of the observed Local Group dSphs, we again apply Equation
\ref{eq:projvirial} to find the values of $M_{200}$ that would give the
observed global velocity dispersions of these systems.  Now, however, we allow
the original NFW halo to be modified either by 1) growing a core of radius
$r_{\rm core}$ with $n_{\rm core}=1$, or 2) steepening of the density profile
beyond radius $r_{\rm trunc}$ to a log-slope of $-\delta=-5$ (all other
parameters in Equation \ref{eq:corenfwtide} are held fixed at NFW values). 
For simplicity, we calculate all models holding the concentration fixed at
$c_{200}=15$, a value typical of low-mass dark matter halos formed in
cosmological simulations \citep{dutton14}. We confirm that alternative choices
in the range $10\leq c_{200}\leq 20$ would not significantly alter the same
behaviour in the middle/right columns of Figure \ref{fig:massconcentration}.

Representing the processes of core formation and tidal truncation,
respectively, the middle and right-hand panels of Figure
\ref{fig:massconcentration} display $M_{200}$ and the corresponding ratios
$\log_{10}[M_*/M_{200}]$, $\log_{10}[R_{\rm h}/r_{200}]$ as functions of
$r_{\rm core}$ and $r_{\rm trunc}$.  We find that, compared to the
unadulterated NFW cases, the observational data can accommodate larger
original halo masses when either there forms a core of radius $r_{\rm core}\ga
3$ kpc $\sim R_{\rm half}$, or when tides steepen the density profile beyond
radii $r_{\rm trunc}\la 1$ kpc. However, Figure \ref{fig:massconcentration}
also indicates that there are problems with both scenarios.  The values of
$r_{\rm core}$ that would be sufficiently large to give each object a `normal'
half-light to halo radius ratio correspond to $M_{200}$ values so large that
the ratio of stellar mass to halo mass plummets, rendering the process of core
formation energetically implausible \citep[see][]{Penarrubia2012,dicintio14}. 
Moreover, values of $r_{\rm trunc}$ that would give normal half-light to halo
radius ratios correspond to implausibly large progenitor masses of $M_{200}\ga
10^{12}M_{\odot}$ for Ant 2 and And XIX, in which case it would have been the
Milky Way and M31 that lost mass to their satellites!  

\subsection{Tidal evolution}

We have found that the properties of Ant 2 are inconsistent with expectations
from an isolated NFW halo and proposed two solutions that somewhat alleviate
the tension: feedback coring the dark matter profile and tidal disruption. To
further investigate these scenarios, we run a series of controlled $N$-body
simulations of a dwarf galaxy in the tidal field of the Milky Way. We use the
method described in \cite{Sanders2018} to setup the initial conditions for a
two-component (dark matter and stars) spherical dwarf galaxy on Ant 2's orbit.
We opt for a fixed time-independent axisymmetric Milky Way potential from
\cite{McMillan2017}\footnotemark, place the dwarf at apocentre $\sim13\,\mathrm{Gyr}$ ago
and integrate forwards. In this model, Ant 2 undergoes $6$ pericentric
passages. The shape of the orbit is very similar to that shown in
Fig.~\ref{fig:orbit} despite the different potential.

\footnotetext{ Note that for a MW-like galaxy the physical mass
  growth within $r\sim100$ kpc is only $~10\%$ in the last $\sim 8$
  Gyr \citep[see fig.  2 in][]{2015ApJ...808...40W}. Given the
  location of Ant~2, the evolution of the MW is thus not likely to be a major
  concern.}

We initially selected a $v_\mathrm{max}=20\,\mathrm{km\,s}^{-1}$,
$c_{200}=15.9$ NFW halo which is hypothesised to be the lowest mass
galaxy-hosting dark matter halo \citep[][]{OkamotoFrenk2009}. The scale radius
of the halo is $r_s=1.45\,\mathrm{kpc}$. Choosing a stellar double-power law
density profile with scale radius $1.45\,\mathrm{kpc}$, outer logarithmic
slope $\beta=5$ and transition $\alpha=2$ and either a core (inner slope
$\gamma=0.1$) or a cusp ($\gamma=1$) produced similar results: the velocity
dispersion fell steadily from $\sim14\,\mathrm{km\,s}^{-1}$ to
$8-10\,\mathrm{km\,s}^{-1}$ while the half-light radius fell by
$\sim30\,\mathrm{percent}$ ending at $\sim1\,\mathrm{kpc}$. Changing the inner
stellar and dark matter slope to a more cored $\mathrm{d}\ln\rho/\mathrm{d}\ln
r=-0.1$ (while retaining the enclosed central mass), we found that the decay
of the velocity dispersion was more rapid (destroying the galaxy on the fifth
pericentric passage), while the half-light radius fell more slowly, but could
also increase between pericentric passages \citep[as shown in
][]{Sanders2018}. We note that these results are largely insensitive to the
adopted stellar mass as the stars are subdominant. We have adopted
$M_\star=(M_\mathrm{DM}(<4/3R_\mathrm{h})/(2000M_\odot))^{5/3}M_\odot$,
choosing a stellar mass-to-light ratio of $2.5$.

\begin{figure}
  \includegraphics[width=\columnwidth]{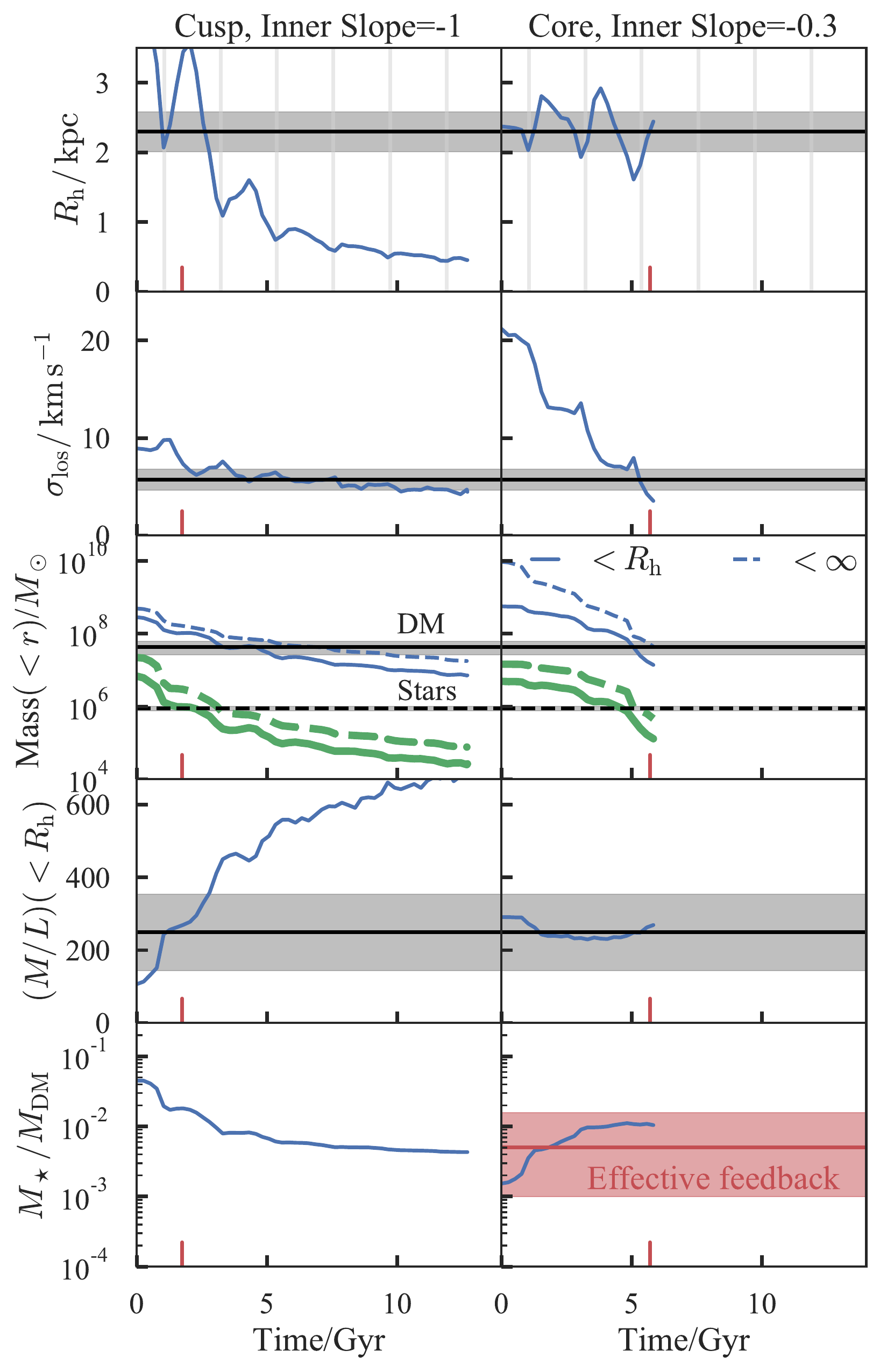}
  \caption{Evolution of two-component $N$-body dwarf galaxy simulations on the
  orbit of Ant 2: the left column corresponds to a cuspy dark-matter
  simulation and the right a cored simulation. The four rows show
  (`circularised') half-light radius, $R_\mathrm{h}=r_\mathrm{h}\sqrt{b/a}$,
  velocity dispersion, mass (dark matter in thin blue, stars in thicker green,
  solid for within $R_\mathrm{h}$ and dashed total), mass-to-light ratio
  (assuming a stellar mass-to-light ratio of $2.5$), and total stellar to dark
  mass ratio. The black horizontal lines and grey shaded regions give the
  median and $1\sigma$ uncertainty for the corresponding measured properties
  of Ant~2. The red shaded region in the bottom panel gives the range of mass
  ratios for which feedback is effective in producing a cored dark matter
  profile \protect\citep{BullockBK2017}. The vertical grey lines show the
  pericentric passages. The red ticks show the times from the two simulations that the observables approximately match Antlia 2.}
\label{fig::tidal_evolution}
\end{figure}

Inspired by these first experiments, we present two scenarios for the
evolution of Ant~2: a cuspy dark matter scenario and cored dark matter
scenario. The results are shown in Fig.~\ref{fig::tidal_evolution}. We compute
the half-light radius from a Plummer fit to bound particles, and the
dispersion is mass-weighted over the entirety of the dwarf (accounting for
perspective effects due to Ant 2's large size). For the cuspy model (left
column of Fig.~\ref{fig::tidal_evolution}), we adopt
$v_\mathrm{max}=16.6\,\mathrm{km\,s}^{-1}$, $c_{200}=15.9$ producing
$r_s=1.2\,\mathrm{kpc}$ to attempt to match $\sigma_\mathrm{los}$, and set the
scale radius of the stars as $r_\star=7.2\,\mathrm{kpc}$ to attempt to match
$R_\mathrm{h}$. Note we are required to set the characteristic stellar radius significantly larger than the dark matter radius, perhaps to an unphysical degree. The dark matter profile is NFW and the stellar profile has
$(\alpha,\beta,\gamma)=(2,5,1)$. We set the velocity anisotropy of the dark
matter as zero and stars as $-0.5$ (tangential bias). Furthermore, we decrease
the mass of stars in the simulation by a factor of $50$ to attempt to match
the final mass-to-light ratio. We see from Fig.~\ref{fig::tidal_evolution}
that the final velocity dispersion and mass-to-light ratio match those
observed. However, the half-light radius for this model rapidly decays as the
model becomes tidally-truncated. Correspondingly, the enclosed masses (both
dark and stellar) fail to match those observed. We see that early in the
evolution (around $3\,\mathrm{Gyr}$, after a single pericentric passage) the
simulation matches all observables well. However, the simulation shows that a
stellar profile that extends significantly beyond the scale radius of the
dominant dark matter mass component is rapidly truncated by tides and such
configurations only last a few orbital periods. In conclusion, a cuspy model
can reproduce the observables but only with a contrived low-mass dark matter
halo and a highly extended stellar profile. Such a configuration is rapidly
destroyed by the Milky Way's tidal field.

\begin{figure*}
  \includegraphics[width=\textwidth]{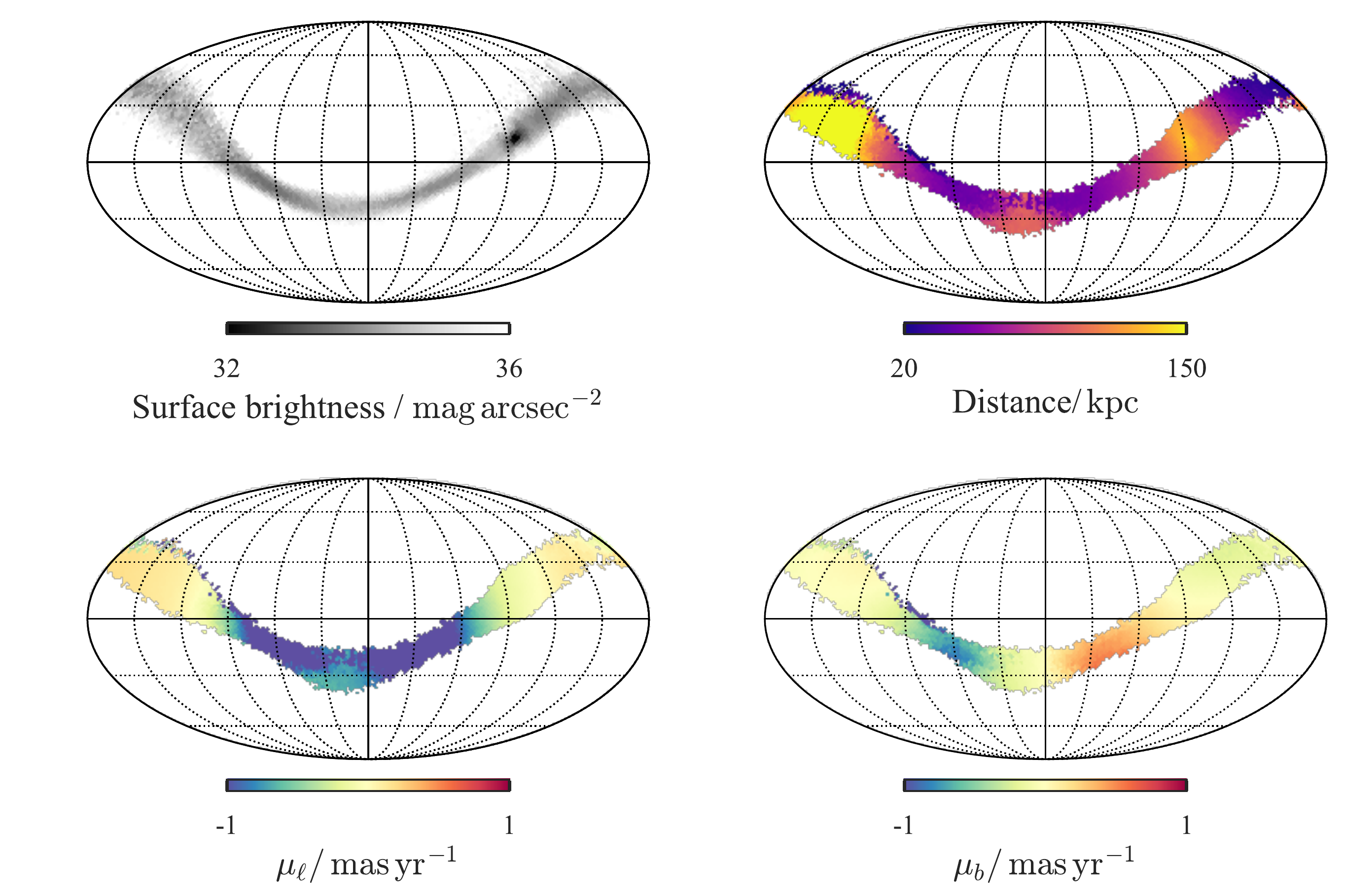}
  \caption{Simulation of Ant~2's disruption on the sky in Galactic coordinates:
  the four panels show the surface brightness, mean distance and galactic
  proper motions for the cored simulation in the left column of
  Fig.~\ref{fig::tidal_evolution}.}
  \label{fig:sim_onsky}
\end{figure*}

As the cuspy model fails to adequately produce all the observables under
reasonable assumptions, we attempt to produce a cored model that
satisfactorily explains the data. We start with
$v_\mathrm{max}=37.4\,\mathrm{km\,s}^{-1}$, $c_{200}=15.9$ producing
$r_s=2.7\,\mathrm{kpc}$ and set the stellar and dark matter inner slope as
$\mathrm{d}\ln\rho/\mathrm{d}\ln r=-0.3$ (retaining the enclosed central mass)
-- this produces a slope of $\mathrm{d}\ln\rho/\mathrm{d}\ln r\approx-0.8$ at
$1.5\,\mathrm{percent}$ of $R_\mathrm{vir}$. We set the scale radius of the
stars as $r_\star=r_s$ (a more reasonable assumption than our over-extended cuspy model) and decrease the mass of stars by a factor of
$\sim250$. We use the same outer slopes, transition slopes and velocity
anisotropies as the cuspy case. From the right column of
Fig.~\ref{fig::tidal_evolution} we find the dispersion falls rapidly, with the
dwarf only surviving three pericentric passages. After the third pericentric
passage the dwarf has a dispersion similar to that of Ant 2. Between
pericentric passages the half-light radius inflates slightly but stays
approximately constant and consistent with the data. The total stellar mass to
dark mass ratio increases by nearly an order of magnitude whilst the mass-to-light ratio stays approximately constant
over the simulation.  There is a compromise between producing an initial total
stellar to dark-matter mass ratio that is high enough to yield cored profiles
via feedback
\citep[$M_\star/M_\mathrm{DM}\sim0.001-0.015$,][]{dicintio14,BullockBK2017}
and a high enough mass-to-light ratio to match that observed for Ant~2.
However, there is some freedom in our choice of stellar mass-to-light ratio,
which was set at $2.5$ but could be larger for a steeper low-mass initial mass
function slope (e.g., Salpeter). Interestingly, the heavy tidal disruption scenario requires an initial stellar mass in the model of
$\sim1\times10^7M_\odot$, placing the progenitor of Ant 2 on the
mass-metallicity relation of the other dwarf galaxies in
Fig.~\ref{fig:massmet}. Furthermore, heavy tidal disruption is also accompanied
by sphericalisation of the dwarf galaxy \citep{Sanders2018} which is possibly
in contradiction with the observed stellar axis ratio of $\sim0.6$. However,
further simulations of flattened progenitors are needed to confirm whether
this is a significant issue. 

\begin{figure}
\includegraphics[width=0.45\textwidth]{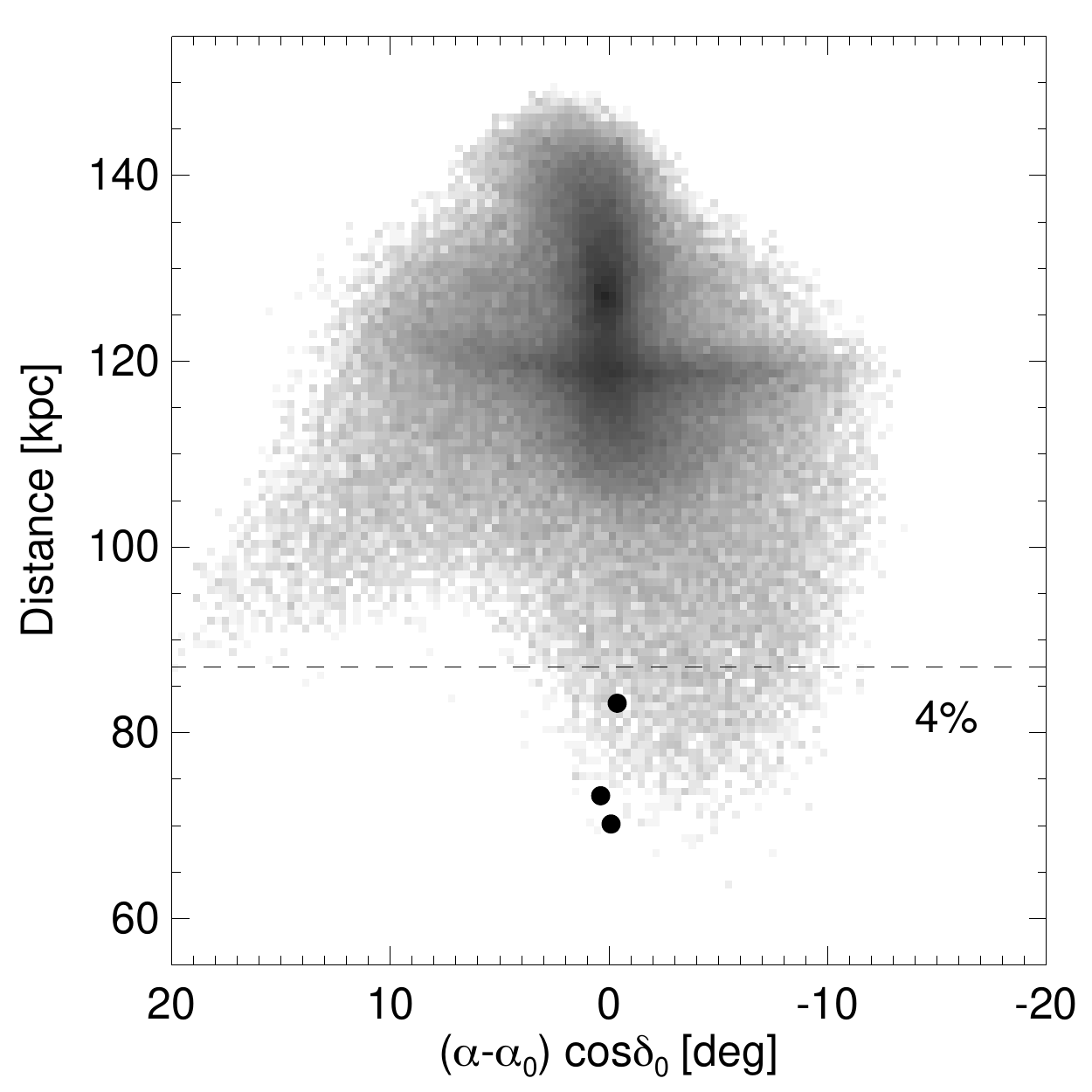}
\caption{Line-of-sight distance distribution of tidal debris at the
  present day for the cored simulation in the right column of
  Fig.~\ref{fig::tidal_evolution} as a function of RA (for the
    same extent in Dec). The dashed line gives the tentative RR Lyrae
  detection limit for \textit{Gaia}. As indicated in the
    Figure, only 4\% of the total number of RR Lyrae in this
    (large) area of the sky are below the dashed line and thus are
    bright enough to be detectable by \textit{Gaia}. The filled black
  circles indicate the locations of the three RR Lyrae with
  $D_h>70\,\mathrm{kpc}$ coincident with the dwarf's location on the
  sky. The horizontal stripes at $D_h\approx120\,\mathrm{kpc}$ and
  $D_h\approx130\,\mathrm{kpc}$ are the ends of the leading and
  trailing debris from the penultimate pericentric passage.}
\label{fig:sim_debris_los}
\end{figure}

A prediction of the cored scenario is that the dwarf
galaxy has deposited $\sim90\,\mathrm{percent}$ of its stellar mass into the
Milky Way halo. In Fig.~\ref{fig:sim_onsky} we show the expected surface
brightness of the resulting material (typically
$34-35\,\mathrm{mag\,arcsec}^{-2}$) as well as the expected median distances
and proper motions using the final apocentric passage from the simulation
rotated to the present coordinates of Ant 2. The line-of-sight distribution of
the tidal debris in this simulation is shown in Fig.~\ref{fig:sim_debris_los}.
Given the current dwarf's location near the apocentre, a substantial portion
of the debris is distributed along the line-of-sight, extending as far as 50
kpc from the progenitor towards the Sun. It appears that, given
\textit{Gaia's} bright limiting magnitude for detecting RR Lyrae stars (dashed
line), only the near side of the debris cloud can be detected using the RR
Lyrae catalogue discussed above (as illustrated by the dashed line and the
filled black circles). If our interpretation is correct, then the three RR
Lyrae detected are only a small fraction of the dwarf's total cohort of such
stars; as noted above, the existence of a much larger undetected population of
RR Lyrae would be consistent with recent results obtained for the Crater~2
dwarf \citep{Joo2018,2018MNRAS.479.4279M}, which is of comparable total
luminosity. Our cored simulation predicts $\sim250$ RR Lyrae within a
$20\times20\,\mathrm{deg}$ box of Ant 2 (assuming Gaia observes all RR Lyrae out to $90\,\mathrm{kpc}$ -- a limiting magnitude of $G=20.27$), whilst the cuspy simulation predicts almost 28,000! At the representative time we have chosen for the cuspy simulation, the tidal disruption is not significant enough to explain the observed foreground RR Lyrae, lending support to the cored picture.

In conclusion, we have found that the only cuspy profiles that adequately
explain the data have both haloes with smaller $v_\mathrm{max}$ than that
required to explain most other dwarf galaxies, and stellar profiles that
extend further than the effective radius of the dark halo
($R_\mathrm{h}/r_s\sim2$). Such an already unlikely configuration does not
survive long within the tidal field of the Milky Way, as the exposed stellar
profile becomes heavily truncated. However, a cored dark matter profile more
naturally explains the data. In this scenario, Ant~2 is embedded initially in
a larger, more massive dark matter halo which naturally produces a broader
stellar distribution. The dwarf is then heavily tidally disrupted by the Milky
Way's tidal field such that, after a few pericentric passages, the velocity
dispersion has fallen to the observed value while the half-light radius is
unchanged. In this picture, Ant~2 has deposited a large fraction of its
stellar mass onto the Milky Way, possibly explaining the foreground RR Lyrae,
and is expected to fully disrupt during the next pericentric passage. 

It is not wholly evident whether the formation of such a large core is
consistent with feedback. Although the slope of the density profile close to
the centre can be reproduced in some feedback models~\citep{To16}, it is
unclear whether such a large core size is really attainable. There remain
other possibilities for the formation of cores in dwarf galaxies, including
the solitonic cores formed in ultralight scalar dark matter particle
theories~\citep{Sc14,Hui2017}. This theory has had some success with
reproducing smaller, but kiloparsec-sized, cores in the classical dwarf
galaxies~\citep{Sc14,Sc16}.Self-interacting dark matter (SIDM), in which the
cross-section of interaction of the DM particles is velocity dependent, can
also produce cores. These are typically of the size of $\sim 1$
kpc~\citep{Tulin}, though they need some fine-tuning of the value of the ratio
of velocity dispersion to DM particle mass. It is an open question whether a
large core can be produced in a low density and low velocity dispersion DM
halo by SIDM theories. Given the multiple populations seen in the extended
metallicity distribution function of Fig.~\ref{fig:fehhist}, the acquisition
of further line of sight velocities will enable the DM density to be mapped
out using the methods of~\citet{Wa11} and \citet{Am12}. This may enable us to
confirm the existence of a core directly from the data, as well as to measure
its size.  This should help resolve whether Ant~2 is completely consistent
with the predictions of feedback, and perhaps even constrain alternative dark
matter theories such as SIDM.

\subsection{Dark matter annihilation}

Like other Milky Way dwarf spheroidal galaxies, Ant~2 is a potential target
for gamma-ray searches for dark matter annihilation
(e.g.~\citet{2005PhR...405..279B,2015PhRvD..91h3535G} and references therein).
In particular, the dwarfs' old stellar populations and lack of gas make them
very unlikely to emit gamma-rays through conventional processes. For dark
matter particles with mass in the GeV to TeV range, searches in dwarfs with
the Fermi Large Area Telescope~\citep{2009ApJ...697.1071A} are among the most
sensitive probes of dark matter annihilation
\citep[e.g.][]{2017ApJ...834..110A}.

Ant~2's small velocity dispersion for its size indicates a low density dark
matter halo (see previous two sub-sections for details), and since the
gamma-ray flux from annihilation is proportional to the square of the dark
matter density, Ant~2 would likely have quite a low signal compared with other
dwarfs. Using the simple estimator in Eq.~8 of \citet{2016PhRvD..93j3512E} and
the properties of Ant~2 in Table~\ref{tab:Properties} we compute the total
flux from dark matter annihilating within $0.5^\circ$ of the centre of Ant~2.
Assuming a dark matter halo with a $\rho \propto r^{-1}$ density
profile\footnote{The $J$ value increases by no more than 0.4~dex when varying
the logarithmic slope of the density profile between $-0.6$ and $-1.4$.} we
find $\log_{10}J(0.5^\circ) /(\mathrm{GeV^2 cm^{-5}}) \approx 16 \pm 0.4$,
where $J$ is proportional to the gamma-ray flux. This is about 3 orders of
magnitude below the $J$ values of the ``top tier'' Milky Way dwarfs where we
expect the strongest annihilation signals \citep[e.g.][]{2015MNRAS.453..849B}.

Although we do not expect a detectable annihilation signal from Ant~2 an
exploration of the Fermi data illustrates a number of issues that will
confront future searches in such large objects. First, Ant~2 may be
significantly extended as seen by Fermi. Dark matter halos are generally less
concentrated than the stars they host and, in contrast with other dwarfs,
Ant~2's stellar half-light angle is already much larger than Fermi's PSF. We
create surface brightness profiles for dark matter emission and convolve them
with Fermi's (energy-dependent) PSF. The size of Ant~2 as seen by Fermi can be
characterized by the angle at which the PSF-convolved surface brightness drops
to half its maximum value. We find the ratio between this angle and the
corresponding angle for a point source of gamma rays. The most point-like
situation we consider is a spherical NFW halo with scale radius twice the
half-light radius\footnote{We use the ``circularized'' half-light radius $r_h
\sqrt{b/a}=2.3\,\mathrm{kpc}$.} and which is truncated beyond the scale
radius. For 1~GeV gamma rays Ant~2 is 50\% more extended than a point source
(80\% at 10~GeV). For less cuspy density profiles the extension is more
pronounced: a modified NFW profile with $\rho \propto r^{-0.5}$ in its inner
parts yields emission at 1~GeV that is 2.4 times more extended than a point
source (4.3 times more at 10~GeV). The shape of Ant~2's halo is quite
uncertain and so a simple search for a gamma-ray point source may not be
optimal. Nevertheless, an eventual detection of spatially extended emission is
quite powerful as it would yield a direct observation of the halo density
profile \citep[e.g.][]{2015ApJ...801...74G}.

We prepare data from 9.4~years of Fermi observations using the procedure
described by \citet{2018MNRAS.tmp.1695K} except that we include energies from
0.5 to 500~GeV and consider the region within a $20^\circ$ square centred on
Ant~2. As in the above study we fit a model to the region using maximum
likelihood \citep[the model includes isotropic and Galactic diffuse emission
and point sources from the Third Fermi Source
Catalogue,][]{2015ApJS..218...23A}. Fitting an additional trial point source
with a flexible energy spectrum at the location of Ant~2 does not improve the
fit ($2\Delta \log L = 0.75$, where $L$ is the Poisson probability of
obtaining the data given the model). We also examine the energy spectra of
events within various-sized apertures centred on Ant~2 as compared to the
background model and find no significant bumps or discrepancies.

As a way of checking for extended emission we rebin the counts and model maps
from the original $0.05^\circ$ pixelisation into coarser maps with
$0.25^\circ$, $0.5^\circ$, $1^\circ$, and $1.5^\circ$ pixels (where the
central pixel is always centred on Ant~2). In each pixel $p$ we construct a
statistic to measure the discrepancy with the best-fitting background model:
$\chi_p^2 = \sum (c_{pi}-m_{pi})^2/m_{pi}$, where $c_{pi}$ and $m_{pi}$ are
the observed and expected numbers of counts in pixel $p$ and energy bin $i$
and the sum is taken over energy bins between 1 and 10~GeV. Generating sky
maps based on $\chi^2_p$ yields no ``hot spots'' centred on Ant~2.

One major caveat is that the Galactic diffuse template provided by the Fermi
Collaboration already absorbs extended excess emission on scales larger than
about $2^\circ$~\citep{2016ApJS..223...26A}, and so it is possible that Ant~2
has already been included in the background model. We examine the morphology
of the diffuse template and do not find any blob-like emission centred on
Ant~2, though this is somewhat difficult due to Ant~2's proximity to the
Galactic plane, a region with numerous diffuse structures and gradients. In
order to try to suppress the very bright emission from the Galactic plane we
re-extract the data but keep only \texttt{PSF3}
events\footnote{\url{https://fermi.gsfc.nasa.gov/ssc/data/analysis/documentation/Cicerone/Cicerone_Data/LAT_DP.html}}.
This subset of data is roughly the quarter of events with the best direction
reconstruction. While this will not help detect extended emission it does
reduce the glare from the Galactic plane. Examining the \texttt{PSF3} counts
maps reveals two potential point sources within the half-light ellipse of
Ant~2, $\Delta \ell =0.3^\circ, \Delta b = -0.3^\circ$ away from the centre of
the dwarf. Fitting point sources at these locations gives an improvement of
$2\Delta \log L=17$. However, this comes at the cost of greatly increased
model complexity and so the resulting significance of these potential sources
($p$ value) is only around the 5\% level. In any case a dark matter
explanation would require a density profile significantly offset from the
stellar distribution.

Despite the lack of a signal from Ant~2, its huge angular size hints at a
possible opportunity for gamma-ray searches using dwarfs. The most powerful
dark matter searches combine observations of multiple dwarfs by assigning
weights to the gamma-ray events from each
dwarf~\citep{2011PhRvL.107x1303G,2015PhRvD..91h3535G}. For the optimal set of
weights, each dwarf contributes to the expected signal to noise in quadrature
as $\mathrm{SNR}^2 = \sum s^2/b$, where the sum runs over energy and spatial
bins, and $s$ and $b$ are the expected number of dark matter photons and
background events detected in each (generally infinitesimal) bin~\citep[see
Eq.~18 of][]{2015PhRvD..91h3535G}. For dwarfs with identical dark matter
halos, as long as the spatial extent of the emission is smaller than the PSF,
a dwarf's contribution to $\mathrm{SNR}^2$ is proportional to $1/D^4$ (the
amplitude of $s$ scales as the total flux $1/D^2$ while the shape of $s$ is
fixed by the PSF). For a homogeneous distribution of dwarfs (number density
$\propto D^2 dD$), this scaling leads to searches dominated by the nearest
dwarfs, e.g. Ursa Major~II, Coma Berenices, and Segue~1. In contrast, for
dwarfs with extended emission $\mathrm{SNR}^2$ scales as $1/D^2$ (in this case
the amplitude of $s$ is fixed but the shape of the surface brightness profile
contracts by a factor of $1/D$). If Ant~2 is the first of many ``missing
giants'' we may eventually find ourselves with an abundance of relatively
distant, spatially extended gamma-ray targets. Since $J \propto
\sigma_v^4/r_h$~\citep{2016PhRvD..93j3512E}, such dwarfs will be especially
important if they have larger velocity dispersions and/or smaller half-light
radii than Ant~2 (as is perhaps suggested by the isolated position of Ant~2 in
Fig.~\ref{fig:mvrh}). It may turn out that this large population becomes as
important for constraining dark matter particles as the handful of ``point
source-like'' dwarfs that are currently the most informative.

\section{Conclusions}\label{sec:conc}

We have presented the discovery of a new dwarf satellite galaxy of the Milky
Way, Antlia~2 (Ant~2).  Originally detected in \textit{Gaia} DR2 data using a
combination of RR Lyrae, proper motions, parallaxes and shallow broad-band
photometry, this new satellite is also confirmed using deeper archival DECam
imaging as well as AAT 2dF+AAOmega follow-up spectroscopy. The CMD of the
Ant~2 dwarf boasts a broad and well-populated RGB as well as a prominent BHB
sequence, which we use as a standard candle. The resulting dwarf's distance
modulus is $m-M=20.6$, which is consistent with the location of the RGB,
giving an independent confirmation of the BHB distance. In addition, there are
possibly 3 associated RRL stars, lying in front of the dwarf along the line of
sight. These likely represent the near side of an extended cloud of tidal
debris originating from Antlia 2. The angular half-light radius of the new
dwarf is $\sim$1.3 degrees, which translates into a gigantic physical size of
$\sim2.8$ kpc, on par with the measurements of the largest satellite of the
MW, the LMC, but with a luminosity some $\sim 4000$ times fainter.

Using $\sim200$ spectroscopically confirmed RGB member stars, we have measured
the dwarf's velocity dispersion to be $\sim5.7$ km s$^{-1}$, which, combined
with the luminosity $M_V=-9$, yields a high mass-to-light ratio of $\sim$300,
typical for a Galactic dwarf. However, given Ant 2's enormous size, the
implied effective DM density is much lower than that of any other dwarf
satellite studied to date. Assuming an NFW density profile, Ant 2 is hosted by
a relatively light DM halo with $M_{200}<10^9 M_{\odot}$ - close to the lowest
mass inferred for the Galactic dwarfs \citep[see][]{Jethwa2018} - which is not
easy to reconcile with its grotesquely bloated appearance. Even if the DM
density deviates from the canonical NFW shape - be it either due to the
effects of stellar feedback or of the Milky Way's tides - bringing the
object's half-light radius in accordance with the rest of the Galactic
satellite population appears difficult.

Nonetheless, a combination of feedback and tides working in concert may
provide a plausible way to explain the observed properties of Ant 2. This
solution requires a substantially more luminous dwarf to be born in a i)
relatively massive and ii) cored DM halo, which subsequently suffers prolific
tidal stripping. We have shown that for such a cored host, the half-light
radius changes little during the disruption but the velocity-dispersion
plummets. A strong prediction of this model is a large amount of tidal debris
left behind by Ant 2, which could be tested by surveying (with either imaging
or spectroscopy) the area around the dwarf. Note however, that even in this
scenario, the structural properties of Ant 2's progenitor remain extreme. For
example, in the size-luminosity plane shown in Figure~\ref{fig:mvrh}, it must
have occupied the empty space below and to the right of Sgr, with $r_h\sim3$
kpc and $M_V\sim-12$.

Given that it currently appears impossible for Ant 2 to be born with a
half-light radius much smaller than currently measured, its position on the
size-luminosity plane (even extrapolated back in time) may imply that dwarf
galaxy formation can proceed at surface brightness (and density) levels
significantly lower than those so far observed. Hence Ant~2 could be the tip
of an iceberg - a population of extremely diffuse Galactic dwarf galaxies even
fainter than the numerous satellites detected in wide-area photometric surveys
over the past two decades. Fortunately, Gaia data -- as illustrated by this
work -- may be the key to testing this hypothesis.

\section*{Acknowledgements}

It is a pleasure to thank thank Andrey Kravtsov and Jorge Pe\~narrubia for
enlightening discussions that helped to improve the quality of this
manuscript, and Chris Lidman for carrying out the spectroscopic service
observations on the AAT. The authors thank Gisella Clementini and Vincenzo
Ripepi for pointing out an error in the RR Lyrae luminosity calculation in an
earlier version of this manuscript. The anonymous referee is thanked for many helpful comments.

This project was developed in part at the 2018 NYC Gaia Sprint, hosted by the
Center for Computational Astrophysics of the Flatiron Institute in New York
City. The research leading to these results has received funding from the
European Research Council under the European Union's Seventh Framework
Programme (FP/2007-2013) / ERC Grant Agreement n. 308024. SK and MGW are
supported by National Science Foundation grant AST-1813881. GT acknowledge
support from the Ministry of Science and Technology grant MOST
105-2112-M-001-028-MY3, and a Career Development Award (to YTL) from Academia
Sinica.

This work presents results from the European Space Agency (ESA) space mission
Gaia. Gaia data are being processed by the Gaia Data Processing and Analysis
Consortium (DPAC). Funding for the DPAC is provided by national institutions,
in particular the institutions participating in the Gaia MultiLateral
Agreement (MLA). The Gaia mission website is https://www.cosmos.esa.int/gaia.
The Gaia archive website is https://archives.esac.esa.int/gaia.


This paper includes data gathered with Anglo-Australian Telescope at Siding
Spring Observatory in Australia.  We acknowledge the traditional owners of the
land on which the AAT stands, the Gamilaraay people, and pay our respects to
elders past and present.

This project used data obtained with the Dark Energy Camera (DECam), which was
constructed by the Dark Energy Survey (DES) collaboration. Funding for the DES
Projects has been provided by the U.S. Department of Energy, the U.S. National
Science Foundation, the Ministry of Science and Education of Spain, the
Science and Technology Facilities Council of the United Kingdom, the Higher
Education Funding Council for England, the National Center for Supercomputing
Applications at the University of Illinois at Urbana-Champaign, the Kavli
Institute of Cosmological Physics at the University of Chicago, the Center for
Cosmology and Astro-Particle Physics at the Ohio State University, the
Mitchell Institute for Fundamental Physics and Astronomy at Texas A\&M
University, Financiadora de Estudos e Projetos, Funda{\c c}{\~a}o Carlos
Chagas Filho de Amparo {\`a} Pesquisa do Estado do Rio de Janeiro, Conselho
Nacional de Desenvolvimento Cient{\'i}fico e Tecnol{\'o}gico and the
Minist{\'e}rio da Ci{\^e}ncia, Tecnologia e Inovac{\~a}o, the Deutsche
Forschungsgemeinschaft, and the Collaborating Institutions in the Dark Energy
Survey. %

The Collaborating Institutions are Argonne National Laboratory, the University
of California at Santa Cruz, the University of Cambridge, Centro de
Investigaciones En{\'e}rgeticas, Medioambientales y Tecnol{\'o}gicas-Madrid,
the University of Chicago, University College London, the DES-Brazil
Consortium, the University of Edinburgh, the Eidgen{\"o}ssische Technische
Hoch\-schule (ETH) Z{\"u}rich, Fermi National Accelerator Laboratory, the
University of Illinois at Urbana-Champaign, the Institut de Ci{\`e}ncies de
l'Espai (IEEC/CSIC), the Institut de F{\'i}sica d'Altes Energies, Lawrence
Berkeley National Laboratory, the Ludwig-Maximilians Universit{\"a}t
M{\"u}nchen and the associated Excellence Cluster Universe, the University of
Michigan, {the} National Optical Astronomy Observatory, the University of
Nottingham, the Ohio State University, the University of Pennsylvania, the
University of Portsmouth, SLAC National Accelerator Laboratory, Stanford
University, the University of Sussex, and Texas A\&M University.

\begin{table*}
    \caption{Results from the spectroscopic modeling}\label{tab:Properties_sp}
    \centering
    \tiny
    \begin{tabular}{@{}lllcccc}
        \hline
id          &ra       &dec      &$rv_h$      &[Fe/H]       & $\log g$   & $\rm{T}_{\rm{eff}}$\\
            & (deg)   & (deg)   & (km/s)     & (dex)       & (dex)      & (K) \\
        \hline
Antlia2\_001 &144.41   &-36.7678 &290.2$\pm$0.7 &-1.33$\pm$0.17 &1.19$\pm$0.29 &4955$\pm$83   \\
Antlia2\_002 &144.0322 &-36.7156 &294.0$\pm$0.9 &-1.12$\pm$0.11 &4.39$\pm$0.23 &4668$\pm$118  \\
Antlia2\_003 &144.5626 &-36.8088 &275.1$\pm$9.8 &-2.39$\pm$0.54 &4.02$\pm$1.37 &4943$\pm$281  \\
Antlia2\_004 &144.2164 &-36.7546 &289.3$\pm$3.4 &-1.34$\pm$0.38 &3.22$\pm$0.88 &4958$\pm$221  \\
Antlia2\_005 &144.6405 &-36.9363 &284.2$\pm$3.9 &-1.44$\pm$0.38 &3.75$\pm$0.86 &4860$\pm$221  \\
Antlia2\_006 &143.8835 &-36.7709 &292.9$\pm$0.8 &-1.1 $\pm$0.12 &1.3 $\pm$0.26 &5075$\pm$57   \\
Antlia2\_007 &144.2962 &-36.8591 &278.3$\pm$4.1 &-1.62$\pm$0.35 &2.7 $\pm$0.93 &4884$\pm$223  \\
Antlia2\_008 &144.8393 &-37.0967 &8.6  $\pm$3.1 &0.23 $\pm$0.32 &5.2 $\pm$0.64 &4629$\pm$185  \\
Antlia2\_009 &144.738  &-37.0794 &20.4 $\pm$3.4 &-0.48$\pm$0.33 &4.98$\pm$0.68 &4698$\pm$331  \\
Antlia2\_010 &144.3932 &-36.9902 &288.5$\pm$1.4 &-1.23$\pm$0.17 &3.91$\pm$0.27 &4428$\pm$125  \\
Antlia2\_011 &144.0758 &-36.8442 &290.2$\pm$1.9 &-1.29$\pm$0.26 &3.88$\pm$0.52 &4711$\pm$260  \\
Antlia2\_012 &144.2665 &-36.9975 &48.4 $\pm$3.9 &-0.22$\pm$0.34 &5.28$\pm$0.54 &5008$\pm$220  \\
Antlia2\_013 &144.6282 &-37.1997 &293.0$\pm$2.5 &-1.2 $\pm$0.29 &3.96$\pm$0.7  &4746$\pm$201  \\
Antlia2\_014 &144.0369 &-36.9445 &288.0$\pm$1.7 &-1.16$\pm$0.25 &3.63$\pm$0.66 &4960$\pm$218  \\
Antlia2\_015 &144.0136 &-36.8662 &287.8$\pm$5.4 &-1.31$\pm$0.28 &4.81$\pm$0.63 &5136$\pm$264  \\
Antlia2\_016 &144.3143 &-37.0689 &292.8$\pm$1.2 &-0.94$\pm$0.13 &3.8 $\pm$0.23 &4287$\pm$83   \\
Antlia2\_017 &144.5795 &-37.2524 &289.2$\pm$2.5 &-1.16$\pm$0.26 &2.95$\pm$0.74 &4926$\pm$196  \\
Antlia2\_018 &144.4075 &-37.1573 &293.1$\pm$1.3 &-1.0 $\pm$0.12 &2.02$\pm$0.61 &5115$\pm$138  \\
Antlia2\_019 &144.256  &-37.0773 &291.0$\pm$1.8 &-1.26$\pm$0.21 &3.84$\pm$0.52 &4701$\pm$202  \\
Antlia2\_020 &144.2549 &-37.1559 &272.2$\pm$6.8 &-1.24$\pm$0.54 &4.58$\pm$1.06 &4865$\pm$340  \\
Antlia2\_021 &144.323  &-37.1384 &291.8$\pm$1.8 &-1.25$\pm$0.25 &2.42$\pm$0.8  &4957$\pm$149  \\
Antlia2\_022 &144.5144 &-37.3139 &85.6 $\pm$2.8 &-0.97$\pm$0.28 &4.16$\pm$0.61 &5124$\pm$231  \\
Antlia2\_023 &144.6594 &-37.4486 &294.1$\pm$2.4 &-1.07$\pm$0.23 &4.29$\pm$0.5  &4723$\pm$198  \\
Antlia2\_024 &144.185  &-37.1481 &283.7$\pm$3.2 &-1.4 $\pm$0.26 &4.54$\pm$0.6  &4688$\pm$249  \\
Antlia2\_025 &144.3197 &-37.2287 &289.4$\pm$0.9 &-0.84$\pm$0.12 &4.28$\pm$0.26 &4827$\pm$136  \\
Antlia2\_026 &144.013  &-36.9676 &300.2$\pm$0.6 &-0.79$\pm$0.09 &3.9 $\pm$0.14 &4408$\pm$73   \\
Antlia2\_027 &144.2995 &-37.2668 &280.1$\pm$3.2 &-1.47$\pm$0.41 &2.55$\pm$1.37 &4974$\pm$239  \\
Antlia2\_028 &144.0498 &-37.0218 &273.0$\pm$6.5 &-0.37$\pm$0.45 &5.19$\pm$0.9  &5046$\pm$320  \\
Antlia2\_029 &144.1539 &-37.1801 &314.7$\pm$0.8 &-0.14$\pm$0.17 &2.81$\pm$0.37 &4034$\pm$103  \\
Antlia2\_030 &144.4285 &-37.4949 &297.3$\pm$2.6 &-1.24$\pm$0.29 &2.62$\pm$0.81 &5024$\pm$196  \\
Antlia2\_031 &144.2236 &-37.302  &281.7$\pm$6.3 &-2.02$\pm$0.64 &3.53$\pm$1.37 &4799$\pm$385  \\
Antlia2\_032 &144.3628 &-37.4769 &285.5$\pm$4.1 &-1.75$\pm$0.44 &2.63$\pm$0.96 &4707$\pm$247  \\
Antlia2\_033 &144.0026 &-37.0409 &292.1$\pm$0.8 &-1.0 $\pm$0.09 &3.65$\pm$0.59 &4355$\pm$246  \\
Antlia2\_034 &144.3108 &-37.4548 &294.8$\pm$0.9 &-1.67$\pm$0.18 &1.36$\pm$0.4  &4937$\pm$102  \\
Antlia2\_035 &144.0247 &-37.1193 &85.8 $\pm$1.5 &-1.58$\pm$0.5  &5.93$\pm$0.32 &3834$\pm$216  \\
Antlia2\_036 &144.1418 &-37.3015 &296.0$\pm$0.7 &-1.14$\pm$0.1  &3.96$\pm$0.14 &4399$\pm$61   \\
Antlia2\_037 &144.081  &-37.1197 &288.4$\pm$2.0 &-0.91$\pm$0.25 &3.51$\pm$0.64 &4856$\pm$202  \\
Antlia2\_038 &144.0853 &-37.259  &295.3$\pm$2.8 &-1.57$\pm$0.31 &3.45$\pm$0.94 &4748$\pm$239  \\
Antlia2\_039 &144.0885 &-37.1639 &288.3$\pm$5.5 &-2.18$\pm$0.49 &2.91$\pm$1.17 &4828$\pm$323  \\
Antlia2\_040 &144.189  &-37.4559 &292.4$\pm$1.2 &-1.39$\pm$0.19 &2.96$\pm$0.55 &4798$\pm$164  \\
Antlia2\_041 &144.1202 &-37.3303 &296.2$\pm$3.5 &-1.44$\pm$0.6  &2.81$\pm$1.14 &4858$\pm$408  \\
Antlia2\_042 &143.9548 &-36.9605 &284.2$\pm$6.5 &-1.16$\pm$0.38 &4.38$\pm$0.85 &4985$\pm$312  \\
Antlia2\_043 &144.1308 &-37.668  &292.0$\pm$2.6 &-1.32$\pm$0.22 &4.32$\pm$0.5  &4589$\pm$226  \\
Antlia2\_044 &143.8916 &-37.1604 &295.6$\pm$2.3 &-1.19$\pm$0.42 &2.09$\pm$1.01 &4889$\pm$201  \\
Antlia2\_045 &143.8625 &-37.1026 &293.8$\pm$1.3 &-1.52$\pm$0.2  &3.67$\pm$0.53 &4654$\pm$150  \\
Antlia2\_046 &143.9378 &-37.3905 &286.4$\pm$4.2 &-1.42$\pm$0.42 &3.64$\pm$0.93 &5240$\pm$285  \\
Antlia2\_047 &143.8282 &-37.132  &218.4$\pm$4.6 &-1.3 $\pm$0.39 &2.93$\pm$1.31 &5144$\pm$296  \\
Antlia2\_048 &144.0119 &-37.7311 &291.2$\pm$2.3 &-1.88$\pm$0.29 &2.82$\pm$0.55 &4634$\pm$245  \\
Antlia2\_049 &143.9227 &-37.4917 &293.3$\pm$3.1 &-1.93$\pm$0.72 &3.83$\pm$1.0  &5108$\pm$537  \\
Antlia2\_050 &143.8137 &-37.1636 &174.1$\pm$3.2 &-1.15$\pm$0.46 &2.69$\pm$1.73 &4682$\pm$294  \\
Antlia2\_051 &143.7889 &-37.101  &300.4$\pm$0.9 &-0.99$\pm$0.13 &2.22$\pm$0.36 &5119$\pm$72   \\
Antlia2\_052 &143.7875 &-37.2203 &290.4$\pm$0.9 &-1.25$\pm$0.17 &1.89$\pm$0.37 &5069$\pm$73   \\
Antlia2\_053 &143.7774 &-37.5246 &303.9$\pm$1.0 &-1.48$\pm$0.19 &2.18$\pm$0.57 &4739$\pm$148  \\
Antlia2\_054 &143.7444 &-37.6849 &273.4$\pm$6.7 &-2.55$\pm$0.57 &3.84$\pm$1.56 &4727$\pm$254  \\
Antlia2\_055 &143.7364 &-37.5761 &282.0$\pm$4.8 &-3.31$\pm$0.67 &2.83$\pm$1.66 &4458$\pm$283  \\
Antlia2\_056 &143.7462 &-37.1284 &292.3$\pm$0.6 &-1.06$\pm$0.14 &1.9 $\pm$0.33 &5036$\pm$68   \\
Antlia2\_057 &143.6452 &-37.6683 &57.8 $\pm$7.2 &-1.03$\pm$0.63 &6.07$\pm$1.09 &4881$\pm$372  \\
Antlia2\_058 &143.5564 &-37.6037 &296.8$\pm$2.0 &-1.27$\pm$0.24 &3.42$\pm$0.62 &4826$\pm$200  \\
Antlia2\_059 &143.6555 &-37.0654 &280.9$\pm$2.0 &-1.08$\pm$0.28 &3.63$\pm$0.75 &4993$\pm$205  \\
Antlia2\_060 &143.5162 &-37.3626 &287.6$\pm$1.5 &-1.26$\pm$0.25 &2.57$\pm$0.45 &5137$\pm$143  \\
Antlia2\_061 &143.865  &-36.8063 &296.1$\pm$3.2 &-1.68$\pm$0.42 &1.68$\pm$0.96 &5025$\pm$278  \\
Antlia2\_062 &143.545  &-37.2742 &293.4$\pm$0.5 &-1.71$\pm$0.04 &1.02$\pm$0.08 &4796$\pm$14   \\
Antlia2\_063 &143.4109 &-37.5214 &51.7 $\pm$4.6 &-0.92$\pm$0.55 &5.82$\pm$1.13 &4574$\pm$344  \\
Antlia2\_064 &143.4977 &-37.3211 &47.7 $\pm$3.0 &-1.06$\pm$0.36 &4.88$\pm$0.67 &4908$\pm$203  \\
Antlia2\_065 &143.3955 &-37.4922 &-24.5$\pm$3.9 &-0.24$\pm$0.35 &5.22$\pm$0.65 &4611$\pm$323  \\
Antlia2\_066 &143.3652 &-37.4772 &294.4$\pm$4.9 &-2.0 $\pm$0.54 &2.55$\pm$1.25 &4821$\pm$331  \\
Antlia2\_067 &143.4774 &-37.1417 &298.4$\pm$0.2 &-0.29$\pm$0.03 &2.49$\pm$0.11 &3995$\pm$17   \\
Antlia2\_068 &143.5907 &-37.0158 &296.2$\pm$7.2 &-1.87$\pm$0.72 &4.42$\pm$1.22 &4856$\pm$416  \\
Antlia2\_069 &143.523  &-37.2347 &300.2$\pm$4.7 &-1.36$\pm$0.56 &3.73$\pm$1.12 &4660$\pm$280  \\
Antlia2\_070 &143.5213 &-37.1668 &290.7$\pm$2.7 &-1.75$\pm$0.33 &3.43$\pm$0.98 &4741$\pm$242  \\
Antlia2\_071 &143.0966 &-37.5431 &289.2$\pm$2.2 &-1.48$\pm$0.27 &2.52$\pm$0.73 &4824$\pm$171  \\
Antlia2\_072 &143.7441 &-36.9139 &29.5 $\pm$3.7 &-1.05$\pm$0.41 &3.71$\pm$1.07 &4971$\pm$328  \\
Antlia2\_073 &143.455  &-37.0521 &287.6$\pm$2.3 &-2.0 $\pm$0.29 &1.71$\pm$0.84 &4928$\pm$206  \\
Antlia2\_074 &143.5503 &-36.9872 &278.5$\pm$2.8 &-1.32$\pm$0.31 &3.36$\pm$0.81 &5178$\pm$230  \\
Antlia2\_075 &143.5367 &-37.0773 &298.8$\pm$3.8 &-1.81$\pm$0.55 &1.76$\pm$0.96 &4852$\pm$285  \\
Antlia2\_076 &143.018  &-37.4241 &294.7$\pm$5.7 &-2.44$\pm$0.64 &4.15$\pm$1.48 &4672$\pm$428  \\
Antlia2\_077 &143.0292 &-37.31   &109.9$\pm$3.6 &-0.83$\pm$0.38 &5.12$\pm$0.84 &4622$\pm$329  \\
Antlia2\_078 &142.894  &-37.3983 &301.4$\pm$3.0 &-1.98$\pm$0.49 &3.22$\pm$0.72 &4859$\pm$253  \\
Antlia2\_079 &143.8168 &-36.7829 &292.7$\pm$2.2 &-1.36$\pm$0.38 &2.64$\pm$0.93 &4793$\pm$212  \\
Antlia2\_080 &142.9567 &-37.3225 &285.2$\pm$0.9 &-2.74$\pm$0.15 &0.67$\pm$0.28 &4492$\pm$73   \\
Antlia2\_081 &143.1255 &-37.2181 &138.6$\pm$3.0 &-1.67$\pm$0.5  &3.25$\pm$0.88 &4967$\pm$225  \\
Antlia2\_082 &142.7703 &-37.3498 &172.1$\pm$2.9 &-1.07$\pm$0.43 &3.59$\pm$0.79 &5080$\pm$208  \\
Antlia2\_083 &143.399  &-36.9386 &281.4$\pm$3.8 &-2.45$\pm$0.5  &2.54$\pm$1.17 &4601$\pm$259  \\
Antlia2\_084 &143.6257 &-36.8654 &293.3$\pm$4.1 &-2.05$\pm$0.47 &4.19$\pm$1.07 &5064$\pm$385  \\
Antlia2\_085 &143.4277 &-36.8772 &295.0$\pm$2.5 &-0.99$\pm$0.45 &2.39$\pm$0.89 &4934$\pm$241  \\
Antlia2\_086 &143.6338 &-36.9415 &285.3$\pm$4.9 &-1.19$\pm$0.53 &2.04$\pm$1.43 &5377$\pm$317  \\
Antlia2\_087 &143.0142 &-37.1148 &11.3 $\pm$2.5 &0.03 $\pm$0.43 &5.72$\pm$0.49 &4549$\pm$275  \\
Antlia2\_088 &142.7531 &-37.2389 &44.2 $\pm$4.7 &-1.97$\pm$0.35 &5.85$\pm$0.33 &4510$\pm$277  \\
Antlia2\_089 &143.6367 &-36.7939 &286.1$\pm$6.3 &-1.46$\pm$0.57 &4.45$\pm$1.07 &4908$\pm$321  \\
Antlia2\_090 &143.1911 &-36.9777 &298.6$\pm$0.9 &-1.42$\pm$0.22 &2.16$\pm$0.66 &4914$\pm$121  \\
Antlia2\_091 &143.7745 &-36.8622 &292.6$\pm$1.7 &-1.21$\pm$0.19 &2.67$\pm$0.53 &5148$\pm$143  \\
Antlia2\_092 &143.5555 &-36.7636 &282.0$\pm$2.7 &-1.58$\pm$0.39 &2.69$\pm$0.83 &4616$\pm$241  \\
Antlia2\_093 &143.5135 &-36.8184 &298.7$\pm$1.2 &-2.04$\pm$0.24 &2.28$\pm$0.51 &4754$\pm$133  \\
Antlia2\_094 &143.3822 &-36.8418 &289.5$\pm$0.5 &-0.96$\pm$0.08 &3.69$\pm$0.12 &4314$\pm$61   \\
Antlia2\_095 &143.0311 &-36.8743 &289.2$\pm$1.2 &-1.25$\pm$0.18 &3.46$\pm$0.41 &4795$\pm$189  \\
Antlia2\_096 &143.248  &-36.8216 &287.8$\pm$1.1 &-2.32$\pm$0.22 &2.13$\pm$0.43 &4677$\pm$139  \\
Antlia2\_097 &143.381  &-36.6321 &301.7$\pm$3.7 &-1.73$\pm$0.41 &3.12$\pm$1.09 &5052$\pm$305  \\
Antlia2\_098 &143.0082 &-36.8738 &195.3$\pm$3.3 &-1.59$\pm$0.4  &3.26$\pm$1.02 &4877$\pm$211  \\
Antlia2\_099 &143.044  &-36.8536 &294.7$\pm$1.2 &-2.28$\pm$0.21 &2.06$\pm$0.37 &4789$\pm$154  \\
Antlia2\_100 &143.2149 &-36.805  &289.8$\pm$1.9 &-1.28$\pm$0.25 &2.69$\pm$0.7  &4827$\pm$190  \\
Antlia2\_101 &143.4385 &-36.641  &296.1$\pm$1.0 &-1.07$\pm$0.19 &2.07$\pm$0.39 &5042$\pm$98   \\
Antlia2\_102 &142.7956 &-36.8848 &31.0 $\pm$3.6 &-1.78$\pm$0.4  &4.35$\pm$0.97 &4470$\pm$300  \\
Antlia2\_103 &142.984  &-36.8579 &3.9  $\pm$1.3 &-0.39$\pm$0.15 &5.35$\pm$0.34 &5123$\pm$196  \\
Antlia2\_104 &143.0315 &-36.7567 &299.1$\pm$3.5 &-2.05$\pm$0.51 &0.01$\pm$1.14 &4899$\pm$337  \\
Antlia2\_105 &143.4262 &-36.7238 &287.9$\pm$1.7 &-1.0 $\pm$0.23 &2.08$\pm$0.66 &5035$\pm$148  \\
Antlia2\_106 &143.485  &-36.6694 &29.4 $\pm$3.3 &0.04 $\pm$0.3  &5.58$\pm$1.01 &4834$\pm$313  \\
Antlia2\_107 &143.567  &-36.6938 &293.5$\pm$3.8 &-1.35$\pm$0.76 &3.04$\pm$1.07 &5391$\pm$461  \\
Antlia2\_108 &142.54   &-36.6496 &69.9 $\pm$6.3 &-1.08$\pm$0.58 &5.89$\pm$0.64 &3966$\pm$619  \\

        \hline
     \end{tabular}
 \end{table*}

\begin{table*}
    \contcaption{}
    \centering
    \tiny
    \begin{tabular}{@{}lllcccc}
        \hline
id          &ra       &dec      &$rv_h$      &[Fe/H]       & $\log g$   & $\rm{T}_{\rm{eff}}$\\
            & (deg)   & (deg)   & (km/s)     & (dex)       & (dex)      & (K) \\
        \hline
Antlia2\_109 &142.48   &-36.6232 &315.4$\pm$5.9 &-0.43$\pm$0.54 &5.47$\pm$0.67 &5021$\pm$199  \\
Antlia2\_110 &142.5243 &-36.6079 &286.2$\pm$9.2 &-1.88$\pm$0.8  &2.88$\pm$1.41 &4869$\pm$239  \\
Antlia2\_111 &143.0022 &-36.6454 &294.2$\pm$3.8 &-2.09$\pm$0.34 &3.28$\pm$0.71 &4860$\pm$228  \\
Antlia2\_112 &143.1721 &-36.6686 &97.5 $\pm$4.2 &-0.45$\pm$0.43 &5.32$\pm$0.86 &4399$\pm$315  \\
Antlia2\_113 &142.8796 &-36.5792 &291.3$\pm$2.0 &-1.53$\pm$0.29 &2.66$\pm$0.61 &4483$\pm$199  \\
Antlia2\_114 &142.715  &-36.5071 &-13.1$\pm$4.4 &-0.71$\pm$0.47 &2.34$\pm$1.91 &4645$\pm$372  \\
Antlia2\_115 &143.0283 &-36.5595 &300.4$\pm$1.2 &-1.52$\pm$0.21 &2.42$\pm$1.01 &4731$\pm$151  \\
Antlia2\_116 &143.3247 &-36.6292 &298.7$\pm$1.2 &-1.47$\pm$0.21 &3.47$\pm$0.51 &4887$\pm$156  \\
Antlia2\_117 &143.0218 &-36.5246 &27.8 $\pm$3.2 &-0.18$\pm$0.32 &5.46$\pm$0.66 &4564$\pm$302  \\
Antlia2\_118 &142.9719 &-36.4776 &299.6$\pm$0.4 &0.33 $\pm$0.21 &1.14$\pm$0.38 &3465$\pm$204  \\
Antlia2\_119 &143.2491 &-36.5633 &293.0$\pm$0.5 &-0.97$\pm$0.09 &3.49$\pm$0.14 &4305$\pm$71   \\
Antlia2\_120 &143.48   &-36.5207 &170.4$\pm$3.3 &-1.27$\pm$0.4  &5.06$\pm$0.73 &5408$\pm$283  \\
Antlia2\_121 &143.1997 &-36.5209 &45.0 $\pm$2.9 &0.25 $\pm$0.3  &4.63$\pm$0.94 &4894$\pm$319  \\
Antlia2\_122 &143.6304 &-36.5916 &292.5$\pm$1.0 &-1.21$\pm$0.17 &2.33$\pm$0.68 &4830$\pm$223  \\
Antlia2\_123 &143.4508 &-36.526  &297.5$\pm$3.7 &-1.83$\pm$0.52 &4.71$\pm$0.9  &4947$\pm$314  \\
Antlia2\_124 &143.4412 &-36.7202 &288.6$\pm$4.8 &-0.54$\pm$0.49 &3.09$\pm$1.23 &5324$\pm$337  \\
Antlia2\_125 &142.679  &-36.2331 &289.1$\pm$1.4 &-1.45$\pm$0.31 &2.31$\pm$0.7  &4687$\pm$187  \\
Antlia2\_126 &142.9998 &-36.3547 &18.2 $\pm$4.9 &-0.45$\pm$0.69 &4.44$\pm$1.4  &4955$\pm$447  \\
Antlia2\_127 &143.1023 &-36.3311 &292.8$\pm$3.5 &-1.62$\pm$0.36 &2.96$\pm$0.92 &4677$\pm$274  \\
Antlia2\_128 &143.3845 &-36.536  &281.1$\pm$1.8 &-1.51$\pm$0.21 &1.61$\pm$0.57 &5014$\pm$178  \\
Antlia2\_129 &142.8157 &-36.2154 &288.6$\pm$5.8 &-1.18$\pm$0.63 &0.98$\pm$1.21 &4827$\pm$310  \\
Antlia2\_130 &143.5783 &-36.6125 &287.5$\pm$0.9 &-1.24$\pm$0.11 &3.87$\pm$0.19 &4392$\pm$85   \\
Antlia2\_131 &143.0325 &-36.2573 &280.0$\pm$1.7 &-1.05$\pm$0.26 &3.39$\pm$0.59 &4978$\pm$254  \\
Antlia2\_132 &143.6435 &-36.5501 &289.6$\pm$1.3 &-1.96$\pm$0.23 &1.44$\pm$0.44 &4766$\pm$113  \\
Antlia2\_133 &143.6242 &-36.4029 &283.3$\pm$2.7 &-2.17$\pm$0.39 &2.52$\pm$0.74 &4633$\pm$234  \\
Antlia2\_134 &143.1325 &-36.2867 &77.2 $\pm$3.7 &0.1  $\pm$0.27 &5.38$\pm$0.52 &5031$\pm$272  \\
Antlia2\_135 &143.5472 &-36.3937 &295.0$\pm$3.9 &-1.71$\pm$0.39 &3.47$\pm$0.97 &4684$\pm$270  \\
Antlia2\_136 &143.029  &-36.1825 &430.5$\pm$7.3 &-1.78$\pm$0.5  &5.02$\pm$0.9  &5020$\pm$322  \\
Antlia2\_137 &142.9986 &-36.1471 &194.0$\pm$5.5 &-1.91$\pm$0.55 &2.76$\pm$1.77 &5326$\pm$359  \\
Antlia2\_138 &143.6291 &-36.4992 &289.2$\pm$3.4 &-2.03$\pm$0.47 &4.13$\pm$0.94 &4535$\pm$312  \\
Antlia2\_139 &143.5753 &-36.4902 &295.3$\pm$1.2 &-1.93$\pm$0.23 &1.09$\pm$0.44 &4803$\pm$123  \\
Antlia2\_140 &143.6256 &-36.5625 &295.7$\pm$2.7 &-0.72$\pm$0.36 &3.56$\pm$1.19 &5032$\pm$260  \\
Antlia2\_141 &143.0544 &-36.0141 &44.3 $\pm$3.6 &-0.48$\pm$0.25 &4.88$\pm$0.7  &4651$\pm$220  \\
Antlia2\_142 &143.1813 &-36.1775 &280.8$\pm$3.6 &-1.14$\pm$0.39 &2.83$\pm$1.59 &4747$\pm$282  \\
Antlia2\_143 &142.9533 &-35.9304 &290.0$\pm$7.1 &-2.72$\pm$0.6  &5.48$\pm$1.54 &3997$\pm$263  \\
Antlia2\_144 &143.2108 &-36.1181 &29.9 $\pm$3.0 &0.11 $\pm$0.24 &5.28$\pm$0.58 &4473$\pm$219  \\
Antlia2\_145 &143.1267 &-35.9899 &288.9$\pm$1.3 &-1.6 $\pm$0.26 &2.23$\pm$0.64 &4791$\pm$135  \\
Antlia2\_146 &143.0852 &-35.9434 &34.0 $\pm$4.4 &-1.79$\pm$0.6  &5.74$\pm$0.42 &4041$\pm$294  \\
Antlia2\_147 &143.2008 &-36.1824 &65.4 $\pm$10.0&-1.88$\pm$0.72 &5.01$\pm$1.17 &4502$\pm$420  \\
Antlia2\_148 &143.1406 &-35.979  &9.7  $\pm$3.2 &-0.91$\pm$0.34 &5.16$\pm$0.81 &4389$\pm$276  \\
Antlia2\_149 &143.278  &-36.148  &290.5$\pm$3.2 &-1.61$\pm$0.41 &2.66$\pm$0.84 &5113$\pm$259  \\
Antlia2\_150 &143.3024 &-36.236  &295.0$\pm$6.6 &-2.25$\pm$0.66 &2.05$\pm$1.53 &4393$\pm$344  \\
Antlia2\_151 &143.2906 &-35.9939 &13.2 $\pm$4.0 &-0.77$\pm$0.58 &5.55$\pm$1.01 &4741$\pm$272  \\
Antlia2\_152 &143.2522 &-35.967  &159.2$\pm$1.5 &-0.86$\pm$0.24 &4.74$\pm$0.49 &5128$\pm$234  \\
Antlia2\_153 &143.487  &-36.2642 &296.8$\pm$2.6 &-2.14$\pm$0.35 &2.5 $\pm$0.68 &4853$\pm$244  \\
Antlia2\_154 &143.6713 &-36.3644 &140.1$\pm$3.2 &-2.11$\pm$0.3  &3.17$\pm$0.94 &4955$\pm$207  \\
Antlia2\_155 &143.2321 &-35.8358 &286.6$\pm$4.1 &-2.1 $\pm$0.53 &4.68$\pm$1.06 &5106$\pm$246  \\
Antlia2\_156 &143.3072 &-35.9445 &158.8$\pm$4.3 &-0.85$\pm$0.65 &3.85$\pm$1.33 &5423$\pm$363  \\
Antlia2\_157 &143.442  &-36.0378 &290.9$\pm$2.4 &-1.36$\pm$0.31 &2.83$\pm$0.75 &4850$\pm$172  \\
Antlia2\_158 &143.6224 &-36.2775 &301.7$\pm$0.7 &-1.25$\pm$0.19 &1.36$\pm$0.34 &4971$\pm$95   \\
Antlia2\_159 &143.4061 &-35.957  &296.2$\pm$2.7 &-2.6 $\pm$0.36 &1.69$\pm$0.74 &4620$\pm$114  \\
Antlia2\_160 &143.5705 &-36.0955 &288.1$\pm$2.3 &-1.72$\pm$0.31 &3.23$\pm$0.72 &4872$\pm$207  \\
Antlia2\_161 &143.5017 &-35.8606 &274.8$\pm$1.9 &-1.87$\pm$0.29 &1.61$\pm$0.78 &4633$\pm$173  \\
Antlia2\_162 &143.5503 &-35.9301 &284.7$\pm$3.1 &-2.04$\pm$0.42 &3.4 $\pm$0.84 &5120$\pm$171  \\
Antlia2\_163 &143.606  &-36.0869 &288.3$\pm$6.0 &-2.51$\pm$0.94 &4.93$\pm$1.12 &4317$\pm$405  \\
Antlia2\_164 &143.5992 &-35.9915 &287.0$\pm$0.9 &-1.22$\pm$0.11 &3.83$\pm$0.73 &4447$\pm$196  \\
Antlia2\_165 &143.6506 &-36.2201 &294.3$\pm$1.4 &-1.39$\pm$0.22 &2.17$\pm$0.6  &5036$\pm$139  \\
Antlia2\_166 &143.7377 &-36.0909 &185.4$\pm$1.6 &-0.26$\pm$0.21 &3.56$\pm$0.62 &5084$\pm$168  \\
Antlia2\_167 &143.6639 &-36.2787 &287.2$\pm$0.6 &-0.99$\pm$0.16 &1.73$\pm$0.72 &4860$\pm$122  \\
Antlia2\_168 &143.6661 &-35.9863 &34.9 $\pm$4.6 &-0.81$\pm$0.51 &5.03$\pm$2.29 &4350$\pm$276  \\
Antlia2\_169 &143.6968 &-35.7993 &287.3$\pm$1.9 &-0.94$\pm$0.24 &2.8 $\pm$0.6  &4891$\pm$171  \\
Antlia2\_170 &143.7152 &-35.8785 &285.4$\pm$4.9 &-2.49$\pm$0.6  &2.45$\pm$1.44 &4596$\pm$339  \\
Antlia2\_171 &143.762  &-35.7853 &284.8$\pm$1.7 &-1.04$\pm$0.21 &2.07$\pm$0.58 &4882$\pm$155  \\
Antlia2\_172 &143.7126 &-36.4737 &294.6$\pm$0.9 &-1.13$\pm$0.16 &1.79$\pm$0.33 &5012$\pm$98   \\
Antlia2\_173 &143.7959 &-36.0037 &288.7$\pm$0.7 &-0.98$\pm$0.11 &1.81$\pm$0.26 &5084$\pm$66   \\
Antlia2\_174 &143.7845 &-36.1298 &287.2$\pm$1.2 &-1.07$\pm$0.17 &3.34$\pm$0.65 &5000$\pm$161  \\
Antlia2\_175 &143.7935 &-36.2443 &27.3 $\pm$2.4 &-0.29$\pm$0.24 &5.08$\pm$0.47 &4873$\pm$180  \\
Antlia2\_176 &143.6763 &-36.3421 &223.7$\pm$2.3 &-1.03$\pm$0.32 &2.69$\pm$0.89 &5169$\pm$195  \\
Antlia2\_177 &143.8092 &-36.2361 &281.9$\pm$3.3 &-1.34$\pm$0.36 &3.83$\pm$0.85 &5254$\pm$262  \\
Antlia2\_178 &143.8774 &-35.9471 &217.0$\pm$3.9 &-1.69$\pm$0.44 &3.02$\pm$0.83 &5170$\pm$226  \\
Antlia2\_179 &143.9661 &-35.7902 &295.7$\pm$6.2 &-1.42$\pm$0.48 &3.94$\pm$1.15 &5203$\pm$288  \\
Antlia2\_180 &143.6692 &-36.3141 &91.8 $\pm$2.4 &0.21 $\pm$0.19 &4.77$\pm$0.52 &4967$\pm$189  \\
Antlia2\_181 &143.9801 &-35.7549 &284.2$\pm$3.2 &-1.28$\pm$0.58 &1.51$\pm$1.48 &4817$\pm$356  \\
Antlia2\_182 &143.8865 &-36.0804 &280.7$\pm$2.4 &-1.87$\pm$0.3  &1.47$\pm$1.07 &4941$\pm$147  \\
Antlia2\_183 &143.834  &-36.2541 &360.4$\pm$3.8 &-0.26$\pm$0.4  &4.84$\pm$0.63 &5785$\pm$335  \\
Antlia2\_184 &143.8395 &-36.3661 &295.2$\pm$1.9 &-1.45$\pm$0.27 &1.8 $\pm$0.56 &4843$\pm$167  \\
Antlia2\_185 &143.6866 &-36.5661 &281.9$\pm$4.2 &-1.41$\pm$0.35 &4.03$\pm$0.88 &4867$\pm$286  \\
Antlia2\_186 &143.942  &-36.2156 &66.4 $\pm$2.8 &0.1  $\pm$0.21 &5.65$\pm$0.38 &5037$\pm$211  \\
Antlia2\_187 &143.6558 &-36.5768 &297.8$\pm$0.6 &-1.0 $\pm$0.09 &1.7 $\pm$0.26 &5098$\pm$50   \\
Antlia2\_188 &144.1966 &-35.7793 &276.5$\pm$5.3 &-1.98$\pm$0.54 &4.07$\pm$1.48 &4682$\pm$266  \\
Antlia2\_189 &144.2014 &-35.8191 &65.2 $\pm$6.6 &-0.86$\pm$0.38 &5.67$\pm$0.49 &4543$\pm$276  \\
Antlia2\_190 &144.0545 &-36.2081 &59.3 $\pm$5.4 &0.04 $\pm$0.37 &5.55$\pm$0.7  &4940$\pm$337  \\
Antlia2\_191 &144.1003 &-36.0622 &378.8$\pm$3.6 &-1.75$\pm$0.52 &2.58$\pm$0.85 &5118$\pm$212  \\
Antlia2\_192 &144.0123 &-36.2622 &295.1$\pm$1.5 &-1.11$\pm$0.17 &2.89$\pm$0.46 &4998$\pm$136  \\
Antlia2\_193 &144.254  &-35.8732 &169.3$\pm$3.4 &-0.63$\pm$0.32 &4.43$\pm$0.98 &5160$\pm$215  \\
Antlia2\_194 &144.1068 &-36.1388 &288.5$\pm$1.2 &-1.85$\pm$0.16 &0.97$\pm$0.4  &4844$\pm$100  \\
Antlia2\_195 &144.239  &-36.0789 &38.8 $\pm$1.0 &0.17 $\pm$0.12 &4.74$\pm$0.28 &4553$\pm$98   \\
Antlia2\_196 &144.2908 &-36.0961 &95.7 $\pm$3.0 &-0.14$\pm$0.23 &5.06$\pm$0.55 &4501$\pm$218  \\
Antlia2\_197 &144.3403 &-35.9994 &107.3$\pm$2.4 &-0.29$\pm$0.28 &4.45$\pm$0.51 &4707$\pm$235  \\
Antlia2\_198 &144.307  &-36.1008 &289.2$\pm$1.2 &-2.06$\pm$0.21 &1.53$\pm$0.45 &4699$\pm$133  \\
Antlia2\_199 &144.1083 &-36.2986 &226.1$\pm$0.7 &-0.68$\pm$0.1  &4.18$\pm$0.18 &4764$\pm$103  \\
Antlia2\_200 &144.0819 &-36.3793 &275.6$\pm$4.6 &-0.95$\pm$0.39 &3.4 $\pm$1.0  &5128$\pm$241  \\
Antlia2\_201 &144.3462 &-36.2044 &281.7$\pm$2.9 &-1.46$\pm$0.3  &4.38$\pm$0.77 &4971$\pm$234  \\
Antlia2\_202 &144.5744 &-36.0101 &19.0 $\pm$4.5 &-0.93$\pm$0.85 &5.03$\pm$0.87 &4963$\pm$268  \\
Antlia2\_203 &144.0653 &-36.4794 &295.3$\pm$0.4 &-1.16$\pm$0.12 &1.26$\pm$0.24 &4940$\pm$46   \\
Antlia2\_204 &144.2034 &-36.365  &297.9$\pm$2.2 &-1.33$\pm$0.3  &3.19$\pm$0.85 &4756$\pm$234  \\
Antlia2\_205 &143.9453 &-36.7579 &320.6$\pm$4.6 &-1.05$\pm$0.45 &3.34$\pm$1.46 &5112$\pm$243  \\
Antlia2\_206 &144.3255 &-36.3558 &290.4$\pm$0.5 &-1.07$\pm$0.06 &1.12$\pm$0.11 &4985$\pm$33   \\
Antlia2\_207 &144.1495 &-36.6055 &297.7$\pm$6.3 &-0.96$\pm$0.52 &4.59$\pm$1.04 &4834$\pm$348  \\
Antlia2\_208 &144.2508 &-36.4576 &65.5 $\pm$2.2 &-0.24$\pm$0.19 &5.47$\pm$0.43 &4750$\pm$149  \\
Antlia2\_209 &144.2995 &-36.4365 &279.2$\pm$0.6 &-1.05$\pm$0.09 &4.01$\pm$0.14 &4397$\pm$76   \\
Antlia2\_210 &144.2863 &-36.5128 &287.3$\pm$0.9 &-0.61$\pm$0.12 &3.77$\pm$0.23 &4679$\pm$119  \\
Antlia2\_211 &144.4146 &-36.4728 &46.9 $\pm$1.3 &0.15 $\pm$0.12 &5.49$\pm$0.28 &4890$\pm$106  \\
Antlia2\_212 &144.267  &-36.5383 &293.0$\pm$1.2 &-1.27$\pm$0.17 &2.95$\pm$0.64 &4904$\pm$114  \\
Antlia2\_213 &144.2461 &-36.5839 &65.3 $\pm$6.3 &-0.28$\pm$0.48 &4.52$\pm$1.08 &4193$\pm$307  \\
Antlia2\_214 &144.3061 &-36.5282 &299.8$\pm$5.0 &-1.49$\pm$0.35 &3.23$\pm$1.06 &4915$\pm$295  \\
Antlia2\_215 &144.125  &-36.7496 &11.3 $\pm$3.5 &-0.04$\pm$0.25 &5.65$\pm$0.37 &4672$\pm$161  \\
Antlia2\_216 &144.1626 &-36.6182 &-5.3 $\pm$1.5 &-0.37$\pm$0.18 &4.16$\pm$0.41 &4906$\pm$151  \\
Antlia2\_217 &144.8575 &-36.4593 &300.9$\pm$1.1 &-0.72$\pm$0.14 &4.16$\pm$0.24 &4377$\pm$86   \\
Antlia2\_218 &144.3082 &-36.6104 &290.4$\pm$2.3 &-1.02$\pm$0.2  &0.08$\pm$0.43 &5241$\pm$182  \\
Antlia2\_219 &144.0521 &-36.7465 &293.0$\pm$7.7 &-1.1 $\pm$0.77 &3.83$\pm$2.16 &5198$\pm$386  \\
Antlia2\_220 &144.4467 &-36.6607 &284.1$\pm$3.1 &-2.33$\pm$0.76 &3.73$\pm$1.05 &5019$\pm$287  \\
Antlia2\_221 &144.4118 &-36.6823 &292.8$\pm$0.8 &-1.02$\pm$0.15 &1.58$\pm$0.66 &4961$\pm$126  \\
        \hline
     \end{tabular}
 \end{table*}






\bibliography{biblio} 







\bsp  
\label{lastpage}
\end{document}